\documentclass[utf8]{frontiersFPHY} 

\usepackage[onehalfspacing]{setspace}

\usepackage{physics}

\usepackage{subcaption}

\usepackage[framemethod=tikz]{mdframed}

\usepackage{pgfplots}
\pgfplotsset{compat=newest}

\usepackage{amsmath}
\usepackage{bm}

\usepackage{graphicx}
\usepackage{rotating}
\usepackage{pifont}
\usepackage{xcolor}
\usepackage{cancel}

\usepackage{scalefnt}
\definecolor{indianred}{rgb}{0.86, 0.08, 0.24}
\definecolor{royalblue}{rgb}{0.25, 0.41, 0.88}
\definecolor{darkorange}{rgb}{1.0, 0.55, 0}
\definecolor{mediumseagreen}{rgb}{0.24, 0.70, 0.44}
\definecolor{purple}{rgb}{0.5, 0, 0.5}
\definecolor{cyan3}{rgb}{0, 0.80, 0.80}

\newcommand{\symboltriangleup}[1][black]{{\color{#1}\scalefont{0.9}{\raisebox{1.5ex}{\begin{turn}{180}$\blacktriangledown$\end{turn}}}}}

\newcommand{\symbolbox}[1][black]{{\color{#1}\scalefont{0.75}{\raisebox{0.2ex}{$\blacksquare$}}}}
\newcommand{\symbolboxopen}[1][black]{{\color{#1}\raisebox{0.2ex}{$\square$}}}
\newcommand{\symbolcircle}[1][black]{{\color{#1}\scalefont{0.75}\raisebox{0.1ex}{\ding{108}}}}

\newcommand{\symboltriangleopen}[1][black]{{\color{#1}$\bm{\bigtriangleup}$}}
\newcommand{\symbolcircleopen}[1][black]{{\color{#1}\scalefont{0.85}\LARGE\raisebox{0.06ex}{$\circ$}}}

\newcommand{\symboldiamondsym}[1][black]{{\color{#1}\scalefont{0.75}\raisebox{-.2ex}{\begin{turn}{45}$\blacksquare$\end{turn}}}}

\newcommand{\symbolstar}[1][black]{{\color{#1}\raisebox{0.2ex}{$\bigstar$}}}

\newcommand{\redcircleopen}{{\scalefont{0.9}\symbolcircleopen[indianred]}}
\newcommand{\redcircle}{{\scalefont{0.9}\symbolcircle[indianred]}}
\newcommand{\bluecircle}{{\scalefont{0.9}\symbolcircle[royalblue]}}
\newcommand{\bluecircleopen}{{\scalefont{0.9}\symbolcircleopen[royalblue]}}

\newcommand{\symbolpentagon}[1][black]{\color{#1}\scalefont{0.75}\raisebox{0.8ex}{\pgfuseplotmark{pentagon*}} } 
\newcommand{\purplepentagon}{{\scalefont{0.9}\symbolpentagon[purple]}}

\newcommand{\bluetriangleupopen}{{\scalefont{0.9}\symboltriangleopen[royalblue]}}
\newcommand{\bluetriangleup}{{\scalefont{0.9}\symboltriangleup[royalblue]}}
\newcommand{\greentriangleup}{{\scalefont{0.9}\symboltriangleup[mediumseagreen]}}
\newcommand{\orangediamond}{{\scalefont{0.9}{\scalefont{0.8}\symboldiamondsym[darkorange]}}}

\newcommand{\redsquare}{{\scalefont{0.9}\symbolbox[indianred]}}
\newcommand{\redsquareopen}{{\scalefont{0.7}\symbolboxopen[indianred]}}

\newcommand{\orangestar}{{\scalefont{0.9}{\scalefont{0.8}\symbolstar[darkorange]}}}

\definecolor{plot1}{rgb}{0.86, 0.08, 0.24}
\definecolor{plot2}{rgb}{0.25, 0.41, 0.88}
\definecolor{plot3}{rgb}{1.0, 0.55, 0}
\definecolor{plot4}{RGB}{61,153,86}

\newcommand{\Nmaxref}{\ensuremath N_\text{max}^\text{ref}}

\def\keyFont{\fontsize{8}{11}\helveticabold }
\def\firstAuthorLast{Tichai \textit{et~al.} } 
\def\Authors{Alexander Tichai\,$^{1,2,3,4,*}$, Robert Roth\,$^{2}$ and Thomas Duguet\,$^{5,6}$}
\def\Address{
$^{1}$ Max-Planck-Institut f\"ur Kernphysik, Heidelberg, Germany \\
$^{2}$ Institut f\"ur Kernphysik, Technische Universit\"at Darmstadt, Darmstadt, Germany \\
$^{3}$ ExtreMe Matter Institute EMMI, GSI Helmholtzzentrum f\"ur Schwerionenforschung GmbH, Darmstadt, Germany \\
$^{4}$ ESNT, Irfu, DPhN, CEA Saclay, Gif-sur-Yvette, France \\
$^{5}$ Irfu, DPhN, CEA Saclay, Gif-sure-Yvette, France \\
$^{6}$ KU Leuven, Instituut voor Kern- en Stralingsfysica, Leuven, Belgium
 }

\def\corrAuthor{
Alexander Tichai\\ 
Institut f\"ur Kernphysik, Technische Universit\"at Darmstadt, Schlossgartenstr. 9, 64289 Darmstadt, Germany
}

\def\corrEmail{alexander.tichai@physik.tu-darmstadt.de}

\newcommand{\Xmax}[1]{ \ensuremath{{#1}_\text{max}} }
\newcommand{\Xmin}[1]{ \ensuremath{{#1}_\text{min}} }

\newcommand{\ra}{\rangle}
\newcommand{\la}{\langle}

\newcommand{\chEFT}{\ensuremath{\chi\text{EFT}}}

\newcommand{\HFB}{|\Phi\ra}
\newcommand{\psiref}{\vert \Psi_\text{ref} \ra }
\newcommand{\psirefbra}{\la \Psi_\text{ref} |}

\newcommand{\primesum}{\sideset{}{'}\sum}

\newcommand{\mcpt}{\text{MCPT}}

\newcommand{\ame}[2]{H^{[4]}_{#1#2}}

\begin{document}
\onecolumn
\firstpage{1}

\title{Many-body perturbation theories for finite nuclei} 

\author[\firstAuthorLast ]{\Authors} 
\address{} 
\correspondance{} 

\extraAuth{}


\newcommand{\remark}[2]
{
\noindent \textbf{#1:} #2
}

\maketitle

\allowdisplaybreaks

\begin{abstract}
In recent years many-body perturbation theory encountered a renaissance in the field of \emph{ab initio} nuclear structure theory. In various applications it was shown that perturbation theory, including novel flavors of it, constitutes a useful tool to describe atomic nuclei, either as a full-fledged many-body approach or as an auxiliary method to support more sophisticated non-perturbative many-body schemes. In this work the current status of many-body perturbation theory in the field of nuclear structure is discussed and novel results are provided that highlight its power as a efficient and yet accurate (pre-processing) approach to systematically investigate medium-mass nuclei. Eventually a new generation of chiral nuclear Hamiltonians is benchmarked using several state-of-the-art flavours of many-body perturbation theory.

\tiny
 \keyFont{ \section{Keywords:} many-body theory, \emph{ab initio}, perturbation theory, correlation expansion, open-shell nuclei}
\end{abstract}

\section{Introduction}

A major goal of quantum many-body theory is to provide accurate solutions of the stationary Schr\"odinger equation
\begin{align}
H |\Psi^{\text{A}}_k\ra = E^{\text{A}}_k | \Psi^{\text{A}}_k\ra \, 
\label{eq:SE}
\end{align}
for a given input Hamiltonian $H$, where $|\Psi^{\text{A}}_k\ra$ denotes the $k$-th eigenstate of the $A$-body system with eigenvalue $E^{\text{A}}_k$. As for the study of the atomic nucleus at low energy, the starting point is a realistic Hamiltonian arising from the modeling of the strong interaction
\begin{align}
H= T + V + W + ... \, , \label{H}
\end{align}
in terms of nucleonic degrees of freedom. In Eq.~\eqref{H}, $T$ denotes the intrinsic kinetic energy, $V$ the two-nucleon (2N) potential, $W$ the three-nucleon (3N) potential and so on. Nowadays, high-precision Hamiltonians are systematically constructed within the framework of chiral effective field theory (\chEFT)~\cite{Weinberg1990,Weinberg1991,Weinberg1992,Entem2003a,Epelbaum2006,Machleidt2011a}. Earlier on, more phenomenological models were employed that fitted a somewhat \emph{ad hoc} parametrization to reproduce few-body observables, e.g. 2N scattering data. Examples of such potentials are the Argonne~\cite{Wiringa1995} or Bonn~\cite{Machleidt2000} potentials. It was recognized in various many-body studies that the inclusion of three-nucleon interactions is mandatory to reproduce nuclear phenomenology, e.g., nuclear saturation properties or the correct prediction of the oxygen neutron dripline~\cite{Otsuka2010,HeBo13}. The relevance of three-nucleon interactions constitutes a key difference from other fields of many-body research like atomic physics, molecular structure, or solid-state physics, and poses significant challenges.

Over the past two decades, the range of applicability of \emph{ab initio} nuclear many-body theory has been extended significantly. While 20 years ago first-principle solutions of the quantum many-body problem were restricted to nuclei lighter than $A\approx 12$, formal and computational developments have made it possible to investigate a much wider range of masses. Initially, the \emph{ab initio} treatment employed mostly large-scale diagonalization methods like the no-core shell model (NCSM)~\cite{NaQu09,RoLa11,BaNa13}, or quantum Monte Carlo (QMC)~\cite{Gezerlis2013,Carlson:2015,Lynn2017,Lynn2019} techniques. In addition,  the few-body ($\text{A}=  3,4$) solution could be constructed using the Fadeev approach or its Fadeev-Yakubowski extension~\cite{Nogga:2001,Lazauskas:2019rfb}. A major breakthrough occurred in the early 2000s when the re-import of so-called \emph{expansion methods} from quantum chemistry provided systematically improvable many-body approximations for medium-mass closed-shell systems up to $^{132}$Sn. In such approaches an initial guess for the exact wave function is taken as a reference state and corrections to this starting point are constructed through a chosen expansion scheme. Initially, this was done within the framework of self-consistent Green's function (SCGF)~\cite{Dickhoff:2004xx,CiBa13,Carbone:2013eqa,SoCi13,Raimondi:2019,Soma:2019bso} and coupled cluster (CC)~\cite{KoDe04,HaPa10,BiLa14,Henderson:2014vka,Jansen:2014,Hagen:2014review} theories that had proven to be extremely efficient at grasping \emph{dynamical correlations} in electronic systems. Later on, the same strategy was transferred to other many-body expansion methods such as many-body perturbation theory (MBPT)~\cite{Roth:2009up,Langhammer2012,Tichai2016,Hu:2016txm,Tichai:2018ncsmpt,Tichai:2018mll}, in-medium similarity renormalization group (IMSRG) ~\cite{Tsukiyama:2011,Tsukiyama:2012,HeBo13,Bo14,H15,Parzuchowski2017,Morris:2017vxi,Stroberg2017} or the unitary model operator approach (UMOA)~\cite{Miyagi2015,Miyagi2017}. All of these methods provide a consistent description of ground-state energies of closed-shell mid-mass nuclei even though the rationales behind their expansions are not trivially related to one another. For sure, this consistency is a remarkable sign of success for \emph{ab initio} nuclear many-body theory.

While closed-shell systems, dominated by so-called \emph{dynamical correlations}, transparently allow for the use of single-reference techniques, the extension to open-shell systems requires a different strategy due to the degeneracy of single Slater-determinant reference states with respect to elementary excitations. Open-shell systems located in the vicinity of shell closures can be targeted via equation-of-motion (EOM) techniques where one or two nucleons are attached to the correlated ground state of a closed-shell nucleus~\cite{PiecuchGour,Jansen:2012ey}. In nuclear systems, however, the strong coupling between spin and orbital angular momenta is such that long sequences of nuclei with open-shell character arise as the mass increases. Consequently, EOM approaches do not provide a viable option to tackle most of the open-shell systems that differ from closed-shell ones by more than one or two mass units. Two different routes have been followed in recent years to overcome this difficulty: i) the construction of \emph{ab initio}-rooted valence-space (VS) interactions used in a subsequent shell-model diagonalization and ii) the use of correlated reference states capturing so-called \emph{static correlations} and thus lifting from the outset the degeneracy with respect to elementary excitations. The design of VS interactions has been performed in various frameworks, going from simple low-order MBPT approaches~\cite{Holt:2014} to more advanced non-perturbative schemes like IMSRG~\cite{Tsukiyama:2012,Stroberg2017} or CC~\cite{Jansen:2014}. While the design of the effective interaction can be performed at low polynomial cost, the final diagonalization, even though taking place in a limited valence space, still exhibits factorial scaling in the number of active nucleons pointing to the \emph{hybrid scaling} of the approach. 

In this article, the construction of VS interactions is not discussed and the focus is rather on the alternative strategy to overcome the limitations of single Slater-determinant-based expansions via the use of more general reference states. Reference states handling the bulk of \emph{static correlations} from the outset re-introduce an \emph{energy gap} in open-shell systems such that expanding the exact ground-state via elementary excitations of the reference state becomes well-defined again. In practice, this is done by resorting to either a \emph{multi-determinantal} reference state or to a single \emph{symmetry-broken} determinantal reference state. In an \emph{ab initio} spirit, this was first done within the frame of SCGF theory, i.e. through the Gorkov extension of SCGF (GSCGF)~\cite{Soma:2011aj} formulated on the basis of a Hartree-Fock-Bogoliubov (HFB) reference state breaking $U(1)$ global-gauge symmetry associated with particle number conservation. Soon after, a multi-reference version of IMSRG (MR-IMSRG) was designed based on a particle-number-projected HFB state~\cite{HergertBogner2014}. Around 2013, these two methods provided the first \emph{ab initio} description of arbitrary mid-mass singly open-shell nuclei. Later on, expansion methods were merged with configuration interaction (CI) technology by using reference states from a prior NCSM calculation performed in a model space of limited size. In this way, one can systematically improve the many-body solution, either by increasing the size of the reference space or by relaxing the truncation of the many-body expansion. Within the framework of perturbative approaches this strategy yields multi-configurational perturbation theory (\mcpt)~\cite{Tichai:2018ncsmpt} whereas for IMSRG it leads to the introduction of the in-medium no-core shell model (IM-NCSM)~\cite{Geb17}.

More recently, the use of particle-number-breaking reference states has been exploited in MBPT and CC theory, thus giving rise to Bogoliubov MBPT (BMBPT) and Bogoliubov CC (BCC). While BCC has only undergone proof-of-principle applications in limited model spaces so far~\cite{Si15}, BMBPT has been applied successfully in large-scale applications up to medium-heavy isotopic chains~\cite{Tichai:2018mll}. Next, the additional or alternative breaking of $SU(2)$ rotational symmetry associated with angular-momentum conservation will provide a systematic access to doubly open-shell nuclei~\cite{Du15}. While intensive efforts are already dedicated to this extension, no systematic result are available yet. 

Whenever using a symmetry-broken reference state, an additional step must be envisioned to account for the quantum fluctuations eventually responsible for the lifting of the fictitious degeneracy associated with the broken symmetry. Indeed, the latter is only {\it emergent} in finite systems~\cite{ui83a,yannouleas07a,Papenbrock:2013cra} such that it is mandatory to restore good symmetry quantum numbers. Doing so does not only change the energy (dramatically in certain situations) but also allows for the proper handling of transition operators characterized by symmetry selection rules. The formalism to achieve this step was proposed recently~\cite{Du15,Duguet:2015yle,qiu17a}, but only applied to a schematic solvable model so far in the nuclear context~\cite{Qiu:2018edx} and will thus not be covered in the present article. Another topic of importance not presently covered relates to the benefit of applying so-called resummation methods to the Taylor expansion associated with perturbation theory. While traditional methods such as Padé resummation~\cite{Baker:1996,Roth:2009up,Langhammer2012} can typically be employed with success, the newly formulated eigenvector continuation method~\cite{Frame:2017fah,Frame:2019jsw} was recently applied successfully to BMBPT~\cite{Demol:2019yjt} and show great promises for the future. Last but not least, and since the present document focuses on finite nuclei, the recent efforts made to describe nuclear matter, at zero and finite temperature, on the basis of MBPT are not discussed~\cite{Hebeler:2011,Drischler2016,Wellenhofer:2018dwh,Drischler2019}.

Eventually, the objective of the present article is to describe the on-going revival of MBPT, both in its basic form and in some of its novel extensions, to describe finite closed-shell and open-shell nuclei. Furthermore, the goal is to demonstrate how MBPT complement non-perturbative methods in two ways, i.e. (i) it can act as an inexpensive pre-processing method to accelerate non-perturbative techniques and (ii) as a post-processing tool to further improve upon non-perturbative many-body approaches.

The present document is structured as follows. Before actually coming to perturbation theory, its possible sources of breakdown are discussed in Sec.~\ref{sec:div}. The nuclear Hamiltonian and softening techniques are introduced in Sec.~\ref{sec:ham}. In Sec.~\ref{sec:MBPT}, formal perturbation theory is laid out. Section~\ref{sec:HFMBPT} is then dedicated to the standard Slater-determinant-based MBPT applicable to closed-shell systems. Multi-configurational perturbation theory and Bogoliubov many-body perturbation theory are discussed as open-shell extensions in Secs.~\ref{sec:mcpt} and \ref{sec:BMBPT}, respectively. In the next two sections, MBPT is employed as a cheap and efficient \emph{pre-processing} tool for non-perturbative many-body methods. In Sec.~\ref{sec:it} MBPT is used to pre-select important configurations, i.e. as a data compression tool, in a non-perturbative calculation whereas in Sec.~\ref{sec:natorbs} it is used to pre-optimize single-particle states, thus accelerating the convergence of the non-perturbative calculation with respect to the one-body basis dimension. In Sec.~\ref{sec:newham}, MBPT is eventually employed as an inexpensive method to provide systematic tests of a newly designed family of $\chEFT$ nuclear Hamiltonians over a large set of nuclei. Conclusions and outlooks are provided in Sec.~\ref{sec:conc}.

\section{Perturbative versus non-perturbative problem}
\label{sec:div}

Before actually discussing MBPT, it is useful to consider the possible reasons for the failure of expansion methods. With these considerations at hand, various MBPT flavors can be better understood based on the interplay between the symmetries characterizing the reference state around which the exact eigenstate is expanded, its single- or multi-determinantal character as well as the resolution scale of the employed Hamiltonian.

\subsection{Rationale}

It has often been argued in the past that the nuclear many-body problem is "intrinsically non-perturbative" such that MBPT is bound to fail to describe correlations between nucleons. The statement is, in such generality, \emph{not true} and the strong disfavour against MBPT techniques is, to a large extent, based on historically grown bias. 

It is crucial to understand the basic fact that the '(non-)perturbative' character of a system characterized by a Hamiltonian $H$ has no meaning in absolute and can only be stated with respect to a chosen starting point. This notion relates, at least implicitly, to an 'unperturbed problem', defined through an unperturbed Hamiltonian $H_0$ and its associated eigenstates, with respect to which the targeted solution is meant to be expanded. If the expansion can be written as a converging powers series in $H_1\equiv H-H_0$, the problem is perturbative. Consequently, the more optimized $H_0$, the better the chances for the problem of actual interest to be {\it made} perturbative. For example, taking free particles as a reference point\footnote{The unperturbed Hamiltonian $H_0$ sums individual kinetic energies in this case.}, the description of any bound state can only be {\it non} perturbative with respect to it. However, more optimized choices of $H_0$, making the description, perturbative might be accessible.

The critical point really resides in the cost required to find an appropriate $H_0$ and its exact eigensolutions, i.e., if the cost to do so is similar to the one needed to employ a non-perturbative method, the perturbation theory built on top of it is not so appealing. In the end, the question is rather: can one find $H_0$ and solve for its eigenstates at a moderate cost such that the eigenstates of $H$ are obtained from them through a converging power series in $H_1\equiv H-H_0$? While success is certainly not guaranteed in general, the search for an optimal, yet simple enough, $H_0$ must be performed in the most open-minded way. Typically, the statement that the nuclear many-body problem is "intrinsically non-perturbative" has been based on too restrictive assumptions of what $H_0$ is allowed to be. 

\subsection{Ultra-violet and infra-red divergences}

\begin{figure}[t!]
\centering
\includegraphics[width=0.5\textwidth]{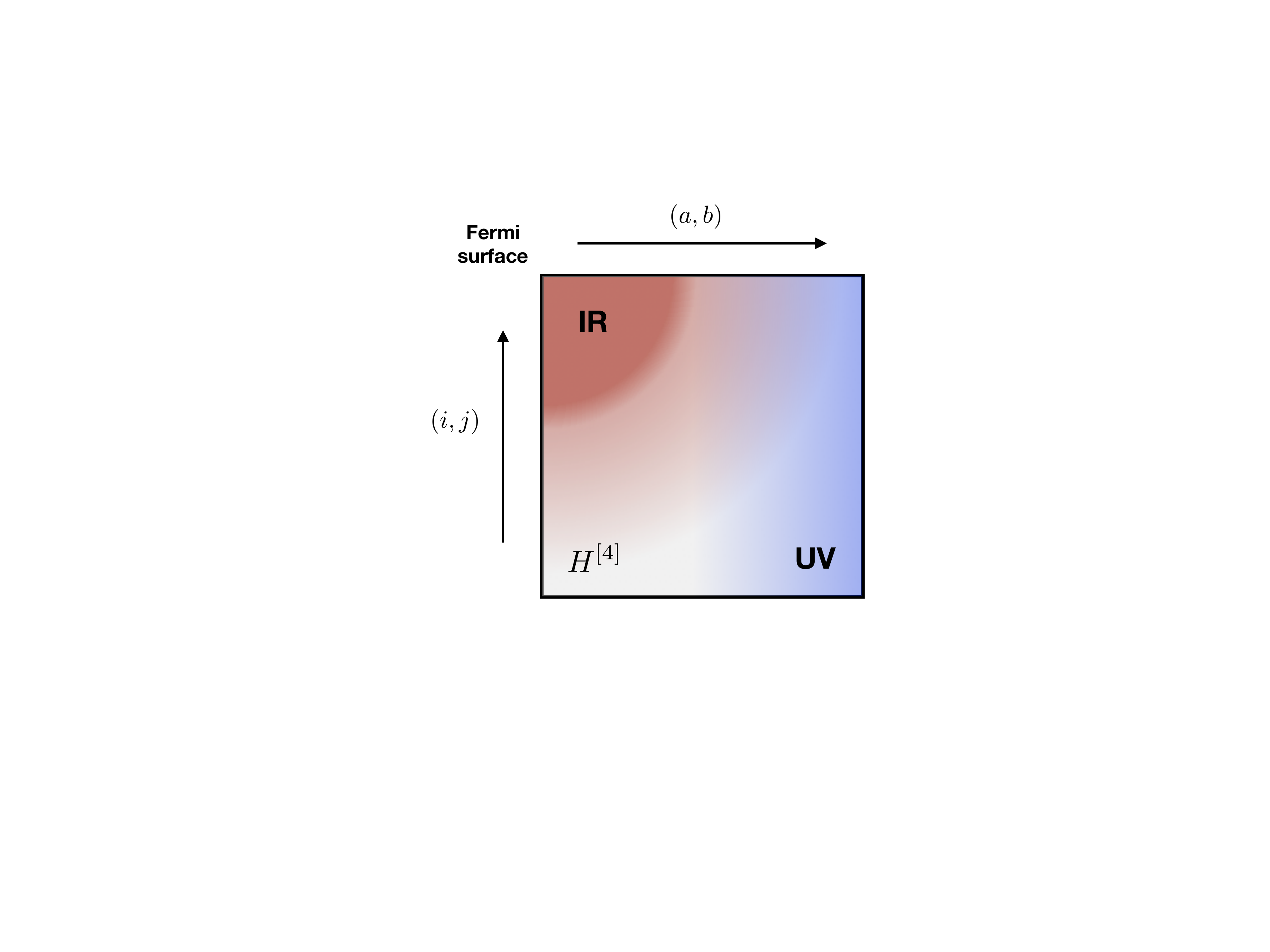}
\caption{Schematic matrix representation of the residual nuclear two-body interaction at play in perturbation theory. The size of the matrix elements is proportional to the darkness of the pixel while the red (blue) color indicate schematically a negative (positive) sign. In the present case, typically associated to MBPT on top of a closed-shell Slater determinant, the IR (UV) couplings translate into large negative (positive) matrix elements between pairs of unoccupied and occupied states near the Fermi level (between any pairs of occupied and highly excited virtual states).}
\label{matrixelementsH1}
\end{figure}

Whether a perturbative approach is viable or not certainly depends on the nature of the Hamiltonian, i.e., on the nature of the elementary degrees of freedom at play and of their interactions. As a matter of fact, two characteristics of the 2N interaction make the many-body problem hard to solve, e.g., \emph{possibly} non-perturbative.  The first one relates to the strong short-range central and tensor-forces between the nucleons inducing strong correlations in the ultra-violet (UV) regime. The second one relates to the large scattering lengths associated with the existence of a weakly bound proton-neutron state and of a virtual di-neutron state, which induce strong many-body correlations in the infra-red (IR) regime. These characteristics induce typical patterns, i.e., large matrix elements, of the residual interaction as schematically depicted in Fig.~\ref{matrixelementsH1}: the upper-left corner corresponds to nuclear matrix elements between high-lying occupied and low-lying virtual states, i.e. to physics around the Fermi surface. Those matrix elements involve strong attractive IR couplings that fall off when moving further away from the Fermi surface. Complementary, repulsive UV couplings become more prominent if energetically higher virtual states are considered, independently of the particular pair of occupied states they interact with.

The potential occurrence of a UV-driven divergence is not nucleus specific and concerns all expansion methods, not just MBPT. Still, specific non-perturbative expansions may, at a given truncation order, resum a specific infinite subset of perturbation theory contributions that appropriately handles the divergence driven by the large UV couplings displayed in Fig.~\ref{matrixelementsH1}. In the end, a pure power series in $H_1$ is certainly the most sensitive expansion to strong low-to-high momentum couplings in the Hamiltonian that typically make the series to diverge after a few reasonable low-order contributions. It happens, however, that UV-driven divergences can be tamed to a large extent via a pre-processing of the Hamiltonian through renormalization group transformations. These transformations are briefly introduced  in Sec.~\ref{sec:srg} and their consequences on the MBPT expansion is illustrated in Sec.~\ref{sec:srgresults}.

\begin{figure}[t!]
\centering
\includegraphics[width=0.70\textwidth]{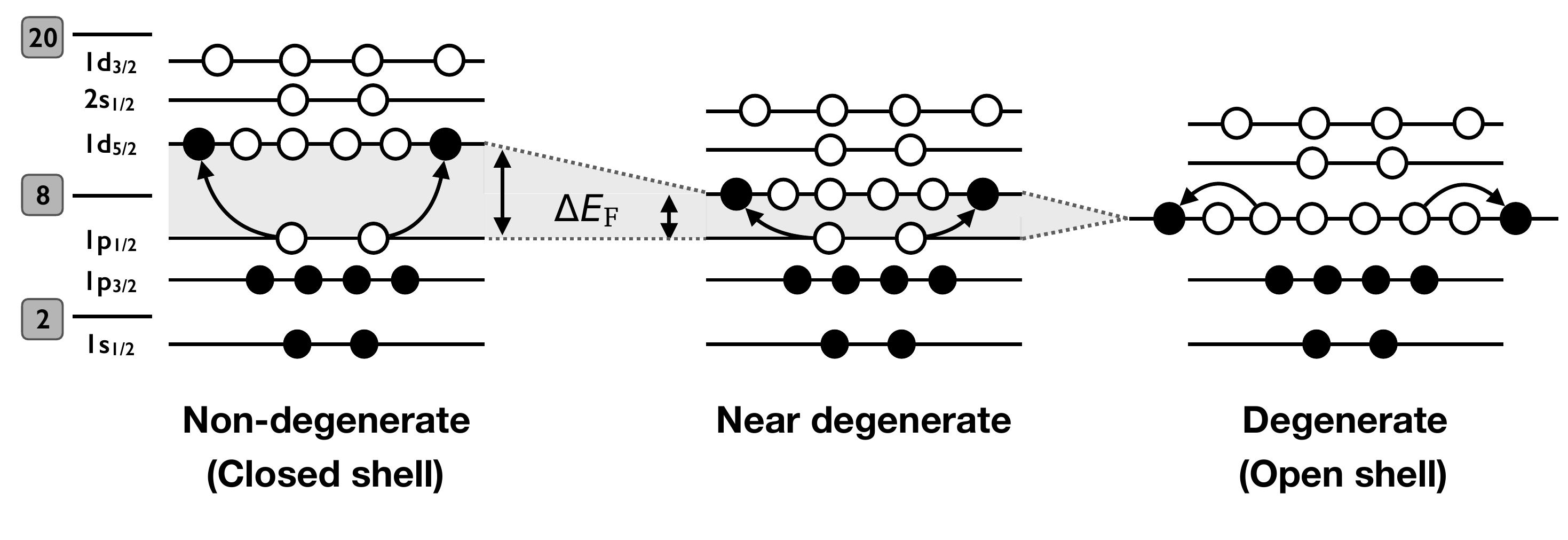}
\includegraphics[width=0.5\textwidth]{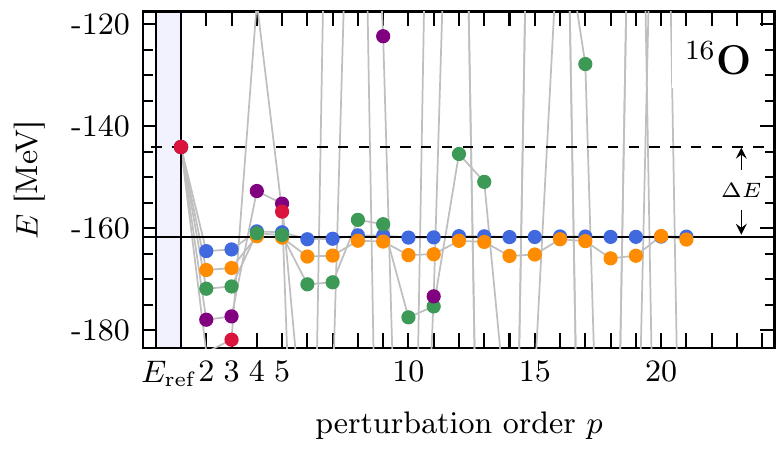}
\caption{\textit{Top panel}: Schematic representation of neutron or proton energy shells and associated occupations corresponding to a two-particle/two-hole excitation on top of the reference Slater determinant, i.e. the ground state of $H_0$, appropriate to a $^{16}$O-like nucleus (N=Z=8). The last occupied shell in the reference state is the \emph{Fermi level} and its energy separation to the first empty level is denoted as $\Delta E_{\text{F}}$. Left: closed-shell nucleus for which the number of nucleons is such that (i) the Fermi level is fully occupied and (ii) $\Delta E_{\text{F}} \gg 0$. Center: sub-closed shell nucleus for which the number of nucleons is such that (i) the Fermi level is fully occupied and (ii) $\Delta E_{\text{F}}$ is small. Right: open-shell nucleus for which the number of nucleons is such that the Fermi level is only partly occupied such that $\Delta E_{\text{F}}=0$.\\
\textit{Bottom panel}: Emergence of an infra-red divergence in the MBPT expansion of the ground-state energy of $^{16}$O induced by a step-wise reduction (going from blue, to yellow, to green, to purple and to red) of the size of the particle-hole gap in the spectrum of $H_0$.}
\label{fig:divergence}
\label{closedvsopenshell}
\label{fig:shelldivcom}
\end{figure}

The potential occurrence of a IR-driven divergence is nucleus specific but concerns all expansion methods, not just MBPT. Infra-red-driven divergences occur whenever the $A$-body unperturbed reference state, i.e. the ground state of $H_0$, is (nearly) degenerate with respect to elementary, e.g. particle-hole, excitations. Considering a standard Slater-determinant reference state, and as illustrated in the top panel of Fig.~\ref{fig:shelldivcom}, this situation occurs whenever the number of constituents is such that the highest occupied shell is only partially filled (is too close to the first empty shell), thus defining so-called open-shell (closed-sub-shell) nuclei. Combined with large IR matrix elements (see Fig.~\ref{matrixelementsH1}), this cancellation (reduction) of the particle-hole gap makes the perturbative expansion (nearly) singular from the outset such that even the few first terms (e.g. Eq.~\eqref{eq:mbpt2}) are not well behaved. A numerical illustration of the emergence of such a IR divergence is provided in the bottom panel of Fig.~\ref{fig:shelldivcom}. This major difficulty can be controlled to a large extent via the use of more general classes of unperturbed Hamiltonians $H_0$ and reference states than typically considered in the past. This idea and the corresponding results are discussed in Secs.~\ref{sec:mcpt} and~\ref{sec:BMBPT}.

Of course, even with the use of renormalization group transformations of the Hamiltonian and rather general classes of reference states, the two characteristics of the nucleon-nucleon interaction in the UV and IR regimes may eventually compromise the convergence of \emph{any} practical perturbative expansion and call for resummation techniques or the use of explicitly non-perturbative methods.

\section{The nuclear Hamiltonian}
\label{sec:ham}

\subsection{The bare operator}

As briefly mentioned in the introduction, chiral effective field theory (\chEFT) provides a convenient framework to construct systematically improvable nuclear Hamiltonians valid in the low-energy regime relevant to nuclear structure~\cite{Entem2003a,Machleidt2011a}.
Starting from nucleons and pions as explicit degrees of freedom, the long- and mid-range parts of the interaction are mediated by multiple-pion exchanges whereas the unresolved short-range part is modelled via contact terms and derivatives of contact terms.
In the early 1990's Weinberg paved the way for a systematic treatment of the strong interaction by introducing a power-counting scheme stipulating the {\it a priori} importance of the infinite number of allowed contributions in the operator expansion~\cite{Weinberg1990,Weinberg1991,Weinberg1992}.
Operators with higher particle rank naturally arise at higher orders in this scheme.
Eventually, the parameters, i.e., the low-energy constants (LEC's), entering the operator expansion are fitted to low-energy experimental data~\cite{Ekstroem2013,Ekstroem2015}. 

Eventually, the second-quantized form of the many-body Hamiltonian takes, in an arbitrary basis $\{| p \rangle \equiv c^{\dagger}_{p}| 0 \rangle\}$ of the one-body Hilbert space  $\mathcal{H}_1$, the form
\begin{align}
H &= T + V + W +\ldots \label{eq:ham} \\
&\equiv  \frac{1}{(1!)^2} \sum _{pq} t_{pq} \, c^{\dagger}_{p} c_{q} \nonumber \\ 
&+ \frac{1}{(2!)^2} \sum _{pqrs} \bar{v}_{pqrs}  \, c^{\dagger}_{p} c^{\dagger}_{q} c_{s} c_{r}   \nonumber \\
&+ \frac{1}{(3!)^2} \sum_{pqrstu} \bar{w}_{pqrstu} \, c^{\dagger}_{p}c^{\dagger}_{q}c^{\dagger}_{r}c_{u}c_{t}c_{s}  \nonumber \\
&+... \, \, .  \nonumber
\end{align}
The Hamiltonian is, thus, represented via a set of one-, two- and three-body matrix elements $t_{pq}$, $\bar{v}_{pqrs}$ and $\bar{w}_{pqrstu}$, respectively. In a modern language the above matrix elements define \emph{tensors} of mode $n=2,4,6$, respectively, where the \emph{mode} specifies the number of indices. 

\subsection{Similarity renormalization group}
\label{sec:srg}

While the tensors defining the Hamiltonian built within $\chEFT$ may display large low-to-high momentum couplings, pre-processing tools can be used to tame them. During the past decade the (free-space) similarity renormalization group (SRG) approach has become the standard technique to generate a ''softened'' basis representation of an operator more amenable to many-body calculations~\cite{Jurgenson2009}.

The SRG approach is based on a unitary transformation of the initial operator $O$ parametrized by a continuous parameter $\alpha \in \mathbb{R}$, i.e.,
\begin{align}
O(\alpha) = U^\dagger(\alpha) O U(\alpha) \, .
\label{eq:srgut}
\end{align}
Equation~\eqref{eq:srgut} can be re-cast into a first-order differential equation
\begin{align}
\frac{d}{d\alpha} O(\alpha) = [\eta(\alpha), O(\alpha)]
\end{align}
involving an anti-Hermitian \emph{generator} $\eta(\alpha)$ that can be chosen freely to achieve a desired decoupling pattern in the transformed operator. A convenient choice employed in many calculations is given by
\begin{align}
\eta(\alpha) \equiv  [ T, O(\alpha) ] \, ,
\end{align}
such that the SRG evolution can be interpreted as a \emph{pre-diagonalization} of the operator in momentum space, thus suppressing the coupling between high- and low-momentum modes. This procedure thus drives the Hamiltonian towards a band-diagonal form. Writing $H(\alpha)\equiv T+V(\alpha)+W(\alpha)+\ldots$ in the same single-particle basis as the starting Hamiltonian, the SRG transformation corresponds to generating $\alpha$-dependent tensors $\bar{v}_{pqrs}(\alpha)$, $\bar{w}_{pqrstu}(\alpha)\ldots$ whose UV elements linking single-particle states corresponding to low and high momenta are strongly suppressed. 

In many-body applications SRG-evolved operators display highly improved model-space convergence, thus facilitating studies of mid-mass nuclei. The impact on the convergence properties of the MBPT series will be illustrated in Sec.~\ref{sec:srgresults}. However, the numerical improvements come at the price of induced many-body operators, i.e., the unitary transformation shifts information to operators with higher particle ranks. For instance, employing an initial two-body operator $O^{2B}$ leads to
\begin{align}
O^{2B} \quad \xrightarrow{\quad\text{SRG}\quad} \quad O^{2B}(\alpha) + O^{3B}(\alpha) + O^{4B}(\alpha) + ... \, .
\label{eq:clustersrg}
\end{align}
In practice, Eq.~\eqref{eq:clustersrg} must be truncated at a given operator rank, thus discarding higher-body operators. This approximation formally violates the unitarity of the transformation in Fock space and eventually induces a dependence of many-body observables on the SRG parameter $\alpha$. A reasonable trade-off must be found for the value of $\alpha$ employed, i.e., it must improve the model-space convergence while keeping the effect of induced many-body operators at a minimum. The optimal parameter range may vary depending on the operator one starts from.

\subsection{The 'standard' Hamiltonian}

All many-body applications discussed below, except for the novel ones presented in Sec.~\ref{sec:newham}, employ a chiral Hamiltonian containing a 2N interaction at next-to-next-to-next-to-leading-order (N3LO) with a cutoff value of $\Lambda_{2\text{N}}= 500\,\text{MeV}/c$~\cite{EnMa03,Navratil2009}. Three-body forces are included up to next-to-next-to-leading order (N2LO)  with a local regulator~\cite{Navratil2009} based on a cutoff value of $\Lambda_{3\text{N}}=400 \,\text{MeV}$~\cite{Roth2012}. This constitutes a 'standard' Hamiltonian used in many recent \emph{ab initio} studies of light and medium-mass nuclei.

Additionally, the intrinsic Hamiltonian is consistently SRG-evolved in the two- and three-body sectors~\cite{Roth2014srg,Maris2014}. The particular value of the SRG parameter is specified in each individual application. To avoid the complication of dealing with genuine three-body operators various forms of so-called normal-ordered two-body approximations (NO2B) are employed, depending on the particular nature of the $A$-body reference state~\cite{Roth2012,Gebrerufael2016,Ripoche2019}.

\section{Formal perturbation theory}
\label{sec:MBPT}

The presentation of perturbation theory can be separated into \emph{formal perturbation theory} and \emph{many-body perturbation theory}~\cite{Shavitt2009}. Formal perturbation theory allows one to understand the general rationale and most relevant properties of the formalism. This is done by employing abstract Dirac notations and by specifying the initial assumptions via the action of Hilbert or Fock space operators on basis vectors. In particular, many key results can be obtained without specifying the content of the Hamiltonian (e.g. the rank of the operators it contains), the nature of the partitioning (e.g. the symmetries characterizing each contribution) and the associated reference state.

\subsection{Partitioning}

The starting point of perturbation theory relates to a \emph{partitioning of the Hamiltonian}
\begin{align}
H \equiv H_0 +  H_1 \, , \label{partitioning}
\end{align}
into an an unperturbed part $H_0$ and a perturbation $H_1 \equiv H- H_0$. The main assumption relies on the fact that the eigenvalue equation for $H_0$ is numerically accessible, i.e.,
\begin{align}
H_0 |\Phi_k \ra = E^{(0)}_k |\Phi_k \ra \, , \label{unperturbed}
\end{align}
delivering the set of unperturbed eigenstates and eigenergies $\{|\Phi_k \ra, E^{(0)}_k ; k \in \mathbb{N}\}$ making up an orthonormal, i.e.
\begin{align}
\la \Phi_k | \Phi_l \ra = \delta_{kl}\, ,
\end{align}
basis of the many-body Hilbert space. 

\remark{Remark}{A large part of this document is dedicated to the description of nuclear ground states, i.e. $k=0$. Consequently, the corresponding index is dropped in the following whenever targeting the ground state, e.g. $| \Psi^{\text{A}}_0 \ra = | \Psi^{\text{A}} \ra$,  $| \Phi_0 \ra = | \Phi \ra$ or $E^{(0)}_0=E^{(0)}$.}

One typically employs \emph{intermediate normalization}, i.e., the ground state $| \Psi^{\text{A}} \ra$ of $H$ is connected\footnote{Both states are supposed to be adiabatically connected when the perturbation $H_1$ is switched on.} to the unperturbed ground-state $| \Phi \ra$ of $H_0$ such that
\begin{align}
1 = \la \Phi | \Psi^{\text{A}} \ra \, .
\end{align}
Associated with the above partitioning are the \emph{projection operators}
\begin{subequations}
\label{space_partitioning}
\begin{align}
P &\equiv |\Phi \ra \la \Phi | \, ,  \\
Q &\equiv 1 - P\, ,
\end{align}
\end{subequations}
where $P |\Phi \ra=|\Phi \ra$ and $Q|\Phi \ra = 0$ by orthonormality. It can be shown that $P$ and $Q$ do meet the requirements of projection operators, i.e., Hermiticity and idempotency~\cite{Shavitt2009}.
The operator $Q$ can be explicitly written as
\begin{align}
Q \equiv \primesum_k |\Phi_k\ra \la \Phi_k | \equiv \sum_{|\Phi_k \ra \neq |\Phi \ra} |\Phi_k\ra \la \Phi_k|  \, ,
\end{align}
where the primed sum indicates the exclusion of the reference state from the summation. With these operators at hand, the exact ground-state can be written as
\begin{align}
|\Psi^{\text{A}} \ra &= P |\Psi^{\text{A}} \ra + Q |\Psi^{\text{A}} \ra \nonumber \\
 &= | \Phi \ra + | \chi \ra\, ,
\label{eq:resid}
\end{align}
where the \emph{correlated part} $| \chi \ra \equiv Q |\Psi^{\text{A}} \ra$, which is the unknown to be solved for, denotes the \emph{orthogonal complement} of $| \Phi \ra$. 

Eventually, the exact ground-state energy is typically accessed in a \emph{projective} way\footnote{The projective character of standard MBPT or CC is to be distinguished from an \emph{expectation value} approach, where the correlated state appears both as the bra and the ket in the evaluation of the energy.} by left-multiplying Eq.~\eqref{eq:SE} with the reference state $\la \Phi |$ such that
\begin{align}
E^{\text{A}} &= \la \Phi | H | \Psi^{\text{A}} \ra \nonumber \\
&= \la \Phi | H_0 | \Phi \ra + \la \Phi | H_1 | \Phi \ra + \la \Phi | H_1 | \chi \ra \nonumber \\
&= E_\text{ref} + \Delta E  \, ,
\label{eq:Eprojective}
\end{align}
where $E_\text{ref} \equiv \la \Phi | H | \Phi \ra = E^{(0)} + \la \Phi | H_1 | \Phi \ra$ and $\Delta E \equiv E^{\text{A}} - E_\text{ref} = \la \Phi | H_1 | \chi \ra$ denote reference and correlation energies, respectively. When using a reference state of product type, e.g. a Slater determinant, $\Delta E$ accounts for correlations between the nucleons beyond the mean-field approximation.

\subsection{Resolvent operator}

The complete derivation of formal perturbation theory is best performed in terms of the (Rayleigh-Schr\"odinger) many-body \emph{resolvent operator}
\begin{align}
R^\text{RS}
\equiv \primesum_k\frac{|\Phi _k \ra \la \Phi_k |}{E^{(0)} - E^{(0)}_k} \, ,
\label{eq:RSres}
\end{align}
which, due to orthonormality of the employed many-body basis, annihilates the reference state
\begin{align}
R^\text{RS} | \Phi \ra = 0 \, .
\end{align}
It is possible to employ alternative choices, such as the \emph{Brillouin-Wigner} resolvent
\begin{align}
R^\text{BW}
\equiv \primesum_k \frac{|\Phi _k \ra \la \Phi_k |}{E^{\text{A}} - E^{(0)}_k} \, ,
\end{align}
which differs from $R^\text{RS}$ by the presence of the \emph{exact energy} in the denominator instead of the unperturbed energy $E^{(0)}$. In practice, Brillouin-Wigner perturbation theory requires an (computationally intensive) iterative solution and, additionally, suffers from a lack of size-extensivity\footnote{A quantum-mechanical method is coined as \emph{size-extensive} if the energy of a systems computed with this method scales linearly in the number of particles.}. Therefore, this choice is only scarcely used in many-body applications. All of the subsequent results are obtained using a Rayleigh-Schr\"odinger resolvent. Consequently, the upper-case label 'RS' is dropped to avoid notational clutter.

\subsection{Power-series expansion}

After a long but straightforward derivation~\cite{Shavitt2009}, one obtains the correlated part of many-body ground-state and associated energy under the form~\cite{Hugenholtz1957,Go57,bloch58a}
\begin{subequations}
\begin{align}
| \chi \ra &= \sum_{k=1}^\infty (R H_1)^{k} | \Phi \ra_c \, , \label{eq:goldstoneS} \\
\Delta E &= \la \Phi | H_1 \sum_{k=1}^\infty (R H_1)^{k} | \Phi \ra_c \, .
\label{eq:goldstoneE}
\end{align}
\end{subequations}
The lower index 'c' stipulates the \emph{connected} character of the expansion ensuring its size-extensivity, i.e., proper scaling of observables with system size\footnote{The first disconnected contribution originally appears at fourth order. It can be shown that the renormalization terms cancel such disconnected contributions at every order~\cite{Go57,bloch58a,Shavitt2009}, thus providing the final connected form.}. Combining Eqs.~\eqref{eq:Eprojective} and~\eqref{eq:goldstoneE}, one obtains the total ground-state wave-function and energy as power series in $H_1$
\begin{align}
| \Psi^{\text{A}} \ra &\equiv \sum_{p=0}^{\infty} |\Psi^{(p)} \ra  \, , \label{seriesS} \\
E^{\text{A}} &\equiv  \sum_{p=0}^{\infty} E^{(p)} \, , \label{seriesE}
\end{align}
such that $|\Psi^{(0)} \ra = |\Phi \ra$ and $E_\text{ref} = E^{(0)} + E^{(1)}$, i.e. the first non-trivial correction contributing to the ground-state correlation energy corresponds to the second-order term of the power series. It reads as
\begin{align}
E^{(2)} = \la \Phi | H_1 R H_1 | \Phi \ra_c \, ,
\end{align}
and can be re-written more explicitly by expanding the resolvent as
\begin{align}
E^{(2)} = \primesum_k \frac{ \la \Phi | H_1 |\Phi_k \ra \la \Phi_k | H_1 | \Phi \ra}{E^{(0)} - E^{(0)}_k} \, .
\label{eq:mbpt2res}
\end{align}
Equation~\eqref{eq:mbpt2res} provides a prototypical example of a PT expression associated with ground-state energy corrections involving a resolvent operator connecting the unperturbed reference state (i.e. the $P$ space) to excited states of $H_0$ (i.e. the $Q$-space), and then going back to the reference state through the perturbation $H_1$. As will be seen with explicit MBPT, the nature of the elementary excitations of the reference state effectively involved at a given order depend on the rank, i.e. the $k$-body character, of the perturbation $H_1$.

\subsection{Recursive formulation}
\label{sec:recursive}

Equations~\eqref{eq:goldstoneS} and \eqref{eq:goldstoneE} conveniently provide explicit expressions for the energy and state corrections whenever working at rather low orders. To go to high orders and study the convergence properties of perturbation theory as a power series, a different scheme becomes more useful. It relates to (i) making more explicit that the perturbative expansion relates to the power series expansion of a mathematical function taken at a particular value of its variable and to (ii) computing the coefficient of the series in a recursive way.

On the basis of the partitioning introduced in Eq.~\eqref{partitioning}, one defines a one-parameter family of Hamiltonians
\begin{align}
H(\lambda) \equiv H_0 + \lambda H_1 \, ,
\end{align}
such that $H(0)=H_0$ and $H(1)=H$. Perturbation theory assumes that the exact eigenstates and eigenenergies of $H(\lambda)$ can be parameterized through the \emph{power series ansatz}
\begin{subequations}
\begin{align}
E(\lambda) &\equiv E^{(0)} + \lambda E^{(1)} +  \lambda^2 E^{(2)} + ... \, , \label{eq:powerseriesE} \\
|\Psi(\lambda)\ra &\equiv |\Psi^{(0)} \ra + \lambda |\Psi^{(1)} \ra +  \lambda^2  |\Psi^{(2)} \ra + ... \, . \label{eq:powerseriesWF}
\end{align}
\label{eq:powerseries}%
\end{subequations}
Setting $\lambda =0$, the problem becomes equivalent to Eq.~\eqref{unperturbed} while setting $\lambda =1$, one recovers the expansions of Eqs.~\eqref{seriesS} and \eqref{seriesE} associated with the fully interacting problem.

Inserting the power series ansatz into the stationary Schr\"odinger equation for $H(\lambda)$ and grouping together the terms proportional to $\lambda^p$ leads to
\begin{align}
E^{(0)} | \Phi \ra + \sum_{p=1}^\infty \lambda^p \Big( H_1 | \Psi^{(p-1)}  \ra + H_0 | \Psi^{(p)}\ra \Big)  = 
E^{(0)} | \Phi \ra + \sum_{p=1} ^\infty \lambda^p \Big( \sum_{j=0}^p E^{(j)} | \Psi^{(p-j)}\ra  \Big) \label{eq:grouped} \, .
\end{align}
Left multiplying Eq. \eqref{eq:grouped} with $\langle \Phi |$ and using 
intermediate normalization  yields
\begin{equation}
\sum_{p=1}^\infty  \lambda^p \langle\Phi | H_1 |
\Psi^{(p-1)}\rangle = \sum_{p=1}^\infty  \lambda^p E^{(p)} \, ,
\end{equation}
which allows one to write the $p$-order ground-state energy correction as
\begin{equation}\label{eq:Ep}
E^{(p)} = \langle\Phi |H_1 | \Psi^{(p-1)}\rangle \, .
\end{equation}
Left multiplying Eq. \eqref{eq:grouped} with $\langle\Phi_m|$, $m\neq
0$ and matching the terms proportional to $\lambda^p$ provides the relation
\begin{equation}\label{eq:Enp}
\left(E^{(0)} - E_m^{(0)}\right) \langle 
\Phi_m |\Psi^{(p)} \rangle    =     \langle \Phi_m| H_1 | 
\Psi^{(p-1)}\rangle  - \sum_{j = 1}^p E^{(j)} \langle 
\Phi_m |\Psi^{(p-j)}\rangle.
\end{equation}

Introducing the coefficients
\begin{equation}\label{eq:Cmnp}
C_{m0}^{(p)} \equiv \langle \Phi_m | \Psi^{(p)} \rangle = 
\frac{1}{E^{(0)} - E_m^{(0)}} 
\left[ \langle \Phi_m| H_1 | \Psi^{(p-1)}\rangle
- \sum_{j = 1}^p E^{(j)} \langle \Phi_m |
\Psi^{(p-j)}\rangle \right],
\end{equation}
allows one to expand the $p$-order ground-state correction $ |\Psi^{(p)} \rangle$ on the unperturbed basis 
$\{|\Phi_m \rangle\}$ according to
\begin{equation}\label{eq:psinp}
| \Psi^{(p)}\rangle = \sum_m C_{m0}^{(p)} |\Phi_m \rangle,
\end{equation}
such that Eq. \eqref{eq:Ep} becomes
\begin{equation}\label{eq:Enp_rec}
E^{(p)} = \sum_m \langle \Phi | H_1 | \Phi_m  \rangle C_{m0}^{(p-1)}.
\end{equation}
Inserting Eq. \eqref{eq:psinp} into Eq. \eqref{eq:Cmnp} further provides a recursive scheme to compute $C_{m0}^{(p)}$ 
\begin{equation} \label{eq:cnmp_rec}
C_{m0}^{(p)} = \frac{1}{E^{(0)} - E_m^{(0)}} 
\left[ \sum_q \langle \Phi_m | H_1 |
\Phi_q \rangle C_{q0}^{(p-1)}
- \sum_{j = 1}^p E^{(j)} C_{m0}^{(p-j)}
\right] \, ,
\end{equation}
with the initial condition $C_{m0}^{(0)}= \delta_{m0}$.

Eventually, Eq.~\eqref{eq:Enp_rec} and \eqref{eq:cnmp_rec} form a set of \emph{recursive relations} from which the ground-state energy and state corrections can be obtained to all orders. In practice, the eigenbasis of $H_0$ and the associated matrix elements of $H_1$  are built such that the latter is stored in memory. The recursive steps are then identified as large-scale matrix-vector multiplications, thus, using the same technology as configuration-interaction approaches like the NCSM. While being formally convenient and obviating the explicit algebraic computation of order-$p$ corrections, this approach is limited by the storage of the Hamiltonian due to the extensive size of the many-body basis. Therefore, only proof-of-principle studies in model spaces of limited dimensionality can be performed and realistic calculations of mid-mass nuclei are out of reach in this way.

\section{Closed-shell many-body perturbation theory}
\label{sec:HFMBPT}

With the results of formal perturbation theory at hand, one can envision to apply them to specific many-body systems. To do so, one must further specify the nature of the partitioning and of the associated reference state. In particular, the maximum rank and symmetries of $H_0$ must be characterized. Additionally, the goal of MBPT is to express all quantities, e.g., many-body matrix elements and unperturbed eigenenergies, entering the formulae at play in terms  of the actual inputs to the many-body problem, i.e. the mode-$2k$ tensors defining the $k$-body contributions to the Hamiltonian (Eq.~\eqref{eq:ham}).

A series of tools exists to compute the expectation value of products of (many) operators in a vacuum state in an incrementally faster, more flexible and less error-prone way. The first step in this series corresponds to using the second-quantized representation of many-body operators in a chosen single-particle basis and to performing canonical commutations of fermionic operators. Next comes Wick's theorem~\cite{wick50a}, which is nothing but a procedure to capture the result in a condensed and systematic fashion. Still, the combinatorics associated with the application of Wick's theorem quickly becomes cumbersome whenever a long string of creation and annihilation operators is involved. Furthermore, many terms thus generated give identical contributions to the end result. Many-body diagrams address this issue~\cite{Shavitt2009} by providing a pictorial representation of the contributions and, even more importantly, by capturing at once all identical contributions, thus reducing the combinatorics tremendously. While being incredibly useful, the number of diagrams itself grows tremendously when applying MBPT beyond the lowest orders, thus leading to yet another combinatorial challenge. This translates into the difficulty to both generate all allowed diagrams at a given order without missing any and to evaluate their expression in a quick and error-safe way. Consequently, the last tool introduced to tackle this difficulty consists of an automatized generation and evaluation of diagrams~\cite{Paldus1973a,Wong1973,Csepes1988,Lyons:1994ew,Kallay2001,Kallay2002,Stevenson2003,Arthuis:2018yoo}. All these technical, yet crucial, aspects of MBPT are not addressed in the present article and the interested reader is referred to the references.

\subsection{Reference state}

The present chapter is dedicated to the simplest form of MBPT appropriate to closed-shell systems. This first version relies on the use of a symmetry-conserving Slater determinant reference state
\begin{align}
|\Phi \ra \equiv \prod_{i=1}^A c^\dagger_i |0\ra \, , \label{refSD}
\end{align}
where the set of single-particle creation operators $\{c^\dagger_p\}$ acts on the physical vacuum $|0\ra$. This constitutes an appropriate starting point of the perturbative expansion as long as $|\Phi \ra $ denotes a closed-shell Slater determinant in agreement with the left-hand case in Fig.~\ref{closedvsopenshell}. While, in principle, the single-particle basis is completely arbitrary, applications will reveal its significant impact on the qualitative behaviour of the perturbative expansion.

\subsection{Normal ordering}

Applying Wick's theorem with respect to $|\Phi\rangle$, the Hamiltonian can be rewritten in terms of normal-ordered contributions
\begin{equation}
\label{hamiltonian normal ordered}
H =  H^{[0]} + \sum_{pq} H^{[2]}_{pq} \, :c^{\dagger}_{p}c_{q}: + \frac{1}{4}\sum_{pqrs} H^{[4]}_{pqrs} \, :c^{\dagger}_{p}c^{\dagger}_{q}c_{s}c_{r}: + \ldots \, ,
\end{equation}
where $::$ denotes the normal order of the involved creation and annihilation operators. Thus, $H^{[0]}$ is the expectation value of $H$ in $|\Phi\rangle$ whereas $H^{[2]}_{pq}$ and $H^{[4]}_{pqrs}$ define matrix elements of {\it effective}, i.e., normal-ordered, one-body and two-body operators, respectively. The dots denote normal-ordered operators of higher ranks, up to the maximum rank $k_{\text{max}}$ characterizing the initial Hamiltonian (Eq.~\eqref{eq:ham}). Through the application of Wick's theorem, an effective operator of rank $k_{\text{eff}}$ receives contributions from all initial operators with rank $k$, where $k_{\text{eff}} \leq k \leq k_{\text{max}}$. Using an initial Hamiltonian with up to three-nucleon interactions and working in the normal-ordered two-body approximation (NO2B)~\cite{Roth2012,Gebrerufael2016}, the residual three-body part $H^{[6]}$ is presently discarded. 
For explicit expressions of the matrix elements defining the normal-ordered operator, see Refs.~\cite{Roth2012,Gebrerufael2016,Ripoche2019}.

\subsection{Partitioning}
\label{sec:partitioning}

To explicitly set up the partitioning of the Hamiltonian (Eq.~\ref{partitioning}), one adds and subtracts a {\it diagonal} normal-ordered one-body operator 
\begin{equation}\label{bar h}
\bar{H}^{[2]} \equiv \sum_{p} e_{p} \, :c^{\dagger}_{p}c_{p}:
\end{equation}
such that
\begin{subequations}
\label{hamiltonian body part}
\begin{align}
H_{0} &= H^{[0]} + \sum_{p} e_{p} \, :c^{\dagger}_{p}c_{p}: \, , \label{hamiltonian body partA}\\
H_{1} &\equiv  \breve{H}^{[2]} + H^{[4]} \, , \label{hamiltonian body partB}
\end{align}
\end{subequations}
with
\begin{equation}\label{breve h}
\breve{H}^{[2]} \equiv H^{[2]} - \bar{H}^{[2]} = \sum_{p \neq q} H^{[2]}_{pq} \, :c^{\dagger}_{p}c_{q}:\, .
\end{equation}

Introducing the set of Slater determinants obtained from $|\Phi\rangle$ via $n$-particle/$n$-hole excitations
\begin{equation}
|\Phi^{ab\cdots}_{ij\cdots}\rangle \equiv c^{\dagger}_{a}c^{\dagger}_{b}\ldots c_{j}c_{i}|\Phi\rangle\, , \label{excitedSD}
\end{equation}
one obtains an orthonormal basis of the $A$-body Hilbert space 
\begin{align}
\mathcal{H}^A = \{ |\Phi \ra, | \Phi^a_i \ra, | \Phi^{ab}_{ij} \ra, | \Phi^{abc}_{ijk} \ra, ... \} \, ,
\end{align}
which is nothing but the eigenbasis of $H_0$
\begin{subequations}
\label{eigenH0}
\begin{align}
H_{0}|\Phi\rangle &= H^{[0]} |\Phi\rangle,\\
H_{0}|\Phi^{ab\cdots}_{ij\cdots}\rangle &= (H^{[0]} +\epsilon^{ab\cdots}_{ij\cdots})|\Phi^{ab\cdots}_{ij\cdots}\rangle\, ,
\end{align}
\end{subequations}
where
\begin{align}
\epsilon^{ab\cdots}_{ij\cdots} &\equiv (e_{a}+e_{b}+\cdots) - (e_{i}+e_{j}+\cdots)\, 
\end{align}
sums (subtracts) the $n$ one-body energies of the particle (hole) states the nucleons are excited into (from). Equation~\eqref{eigenH0} corresponds to the explicit form of Eq.~\eqref{unperturbed} in the case of a Slater determinant reference state. 

\remark{Convention}{
One-body states occupied (unoccupied) in the reference determinant are labelled by $i,j,k,...$ ($a,b,c,...$) and are referred to as \emph{hole} (\emph{particle}) states. Generic one-body states are denoted by $p,q,r,...$.}

The single-particle energies $\{\epsilon_p\}$ are parameters of the theory that are fixed by the partitioning, which itself defines the reference state. They can be chosen arbitrarily as long as the $A$ occupied hole states have lower energies than the remaining particle states, such that $\epsilon^{ab\cdots}_{ij\cdots}>0$. A simple choice employed in nuclear physics consists of building  $|\Phi\rangle$ by filling up the $A$ lowest single-particle eigenstates of the spherical \emph{harmonic oscillator} Hamiltonian~\cite{Roth:2009up,Langhammer2012}, i.e., setting
\begin{align}
H_0 \equiv \frac{\vec{p}^{\,2} }{2m} + \frac{1}{2} m \omega^2 \vec{r}^{\, 2} \, ,
\end{align}
where the oscillator frequency $\omega$ specifies the width of the potential. A more standard choice throughout various fields of many-body physics and chemistry relates to the so-called M{\o}ller-Plesset partitioning that corresponds to taking $\breve{H}^{[2]}=0$, i.e. $H_1 = H^{[4]}$. This is obtained by using the reference Slater determinant $|\Phi\rangle$ solution of the Hartree-Fock (HF) variational problem and by defining $\bar{H}^{[2]}$ from the eigenvalues of the one-body HF Hamiltonian.

\subsection{Perturbative expansion}

Given the eigenbasis of $H_0$ characterized by Eqs.~\eqref{refSD},~\eqref{excitedSD} and~\eqref{eigenH0}, the many-body resolvent (Eq.~\eqref{eq:RSres}) takes the form
\begin{align}
R =
 - \sum_{ai} \frac{|\Phi^a_i\ra \la \Phi^a_i| }{ \epsilon^a_i} -
 \left(\frac{1}{2!}\right)^{2} \sum_{abij} \frac{|\Phi^{ab}_{ij}\ra \la \Phi^{ab}_{ij}| }{ \epsilon^{ab}_{ij}} -
 \left(\frac{1}{3!}\right)^{2}\sum_{abcijk} \frac{|\Phi^{abc}_{ijk}\ra \la \Phi^{abc}_{ijk}| }{ \epsilon^{abc}_{ijk}}+
  ... \, ,
\end{align}
and is to be fed into Eq.~\eqref{eq:goldstoneE} that, once truncated at a given power in $H_1$, provides the correlation energy at the corresponding perturbative order. 

\subsection{Low-order formulas}
\label{sec:mbpt_loworder}

As alluded to above, the evaluation of low-order corrections is facilitated by representing the MBPT expansion diagrammatically. This is typically done using either Hugenholtz or (anti-)symmetrized Goldstone diagrams, i.e. the time-ordered counterpart of Feynman diagrams that are used to compute matrix elements in quantum field theory. The interested reader is referred to the literature, e.g. Ref.~\cite{Shavitt2009}, for an elaborate discussion of the diagrammatic rules and their relation to Wick's theorem.

Focusing on the first non-trivial correction to the reference energy (Eq.~\ref{eq:mbpt2res}), the second-order correction takes the algebraic form
\begin{align}
E^{(2)} = - \sum_{ai} \frac{H^{[2]}_{ai} H^{[2]}_{ia}}{\epsilon^a_i} - \frac{1}{4} \sum_{abij} \frac{ \ame{ab}{ij} \ame{ij}{ab}}{\epsilon^{ab}_{ij}} \, ,
\label{eq:mbpt2}
\end{align}
and is, thus, expressed in terms of the tensors defining the residual interaction $H_1$ (Eq.~\eqref{hamiltonian body partB}) in normal-ordered form. The first contribution in Eq.~\ref{eq:mbpt2} relates to a so-called \emph{non-canonical diagram} that vanishes if the reference state is taken to be the HF Slater determinant. The second term constitutes the genuine and dominant second-order correction that contributes for any Slater determinant reference state. Using the HF Slater determinant reference state has the practical benefit of lowering the number of many-body diagrams to be considered. While this feature is not relevant at second-order, the proliferation of non-canonical diagrams at higher order~\cite{Arthuis:2018yoo} makes the writing of numerical codes more cumbersome. Still, at a given order non-canonical diagrams are always of sub-leading complexity from a computational point of view, i.e. they involve fewer single-particle summations, such that they do not drive the computational cost. 

Since $E^{(2)}$ provides the leading contribution to the perturbative expansion, one observes that dynamical correlations are dominated by low-lying 2p2h-contributions. Most importantly, it is clear from Eq.~\eqref{eq:mbpt2} that the second-order correction is manifestly negative, i.e. it increases the binding energy. This stems from the fact that the numerators are squared norms of matrix elements contributing to $H_1$ and that the denominators are positive as long as the Slater determinant reference state displays a non-zero shell gap between occupied and unoccupied states, i.e., as long as one deals with a closed-shell nucleus.

\subsection{Results}
\label{sec:HFMBPTres}

The first goal of the present analysis is to study the convergence characteristics of the perturbative expansion.  In absence of analytical knowledge, this study must be based on empirical observations of high-order corrections, which is achieved through the recursive formulation of Sec.~\ref{sec:recursive} in small model spaces. Following this analysis, results of low-order MBPT calculations in realistic model spaces are presented to illustrate state-of-the-art \emph{ab initio} applications to doubly closed-shell nuclei~\cite{Tichai2016}.

\begin{figure}[t!]
\centering
\includegraphics[width=0.75\textwidth]{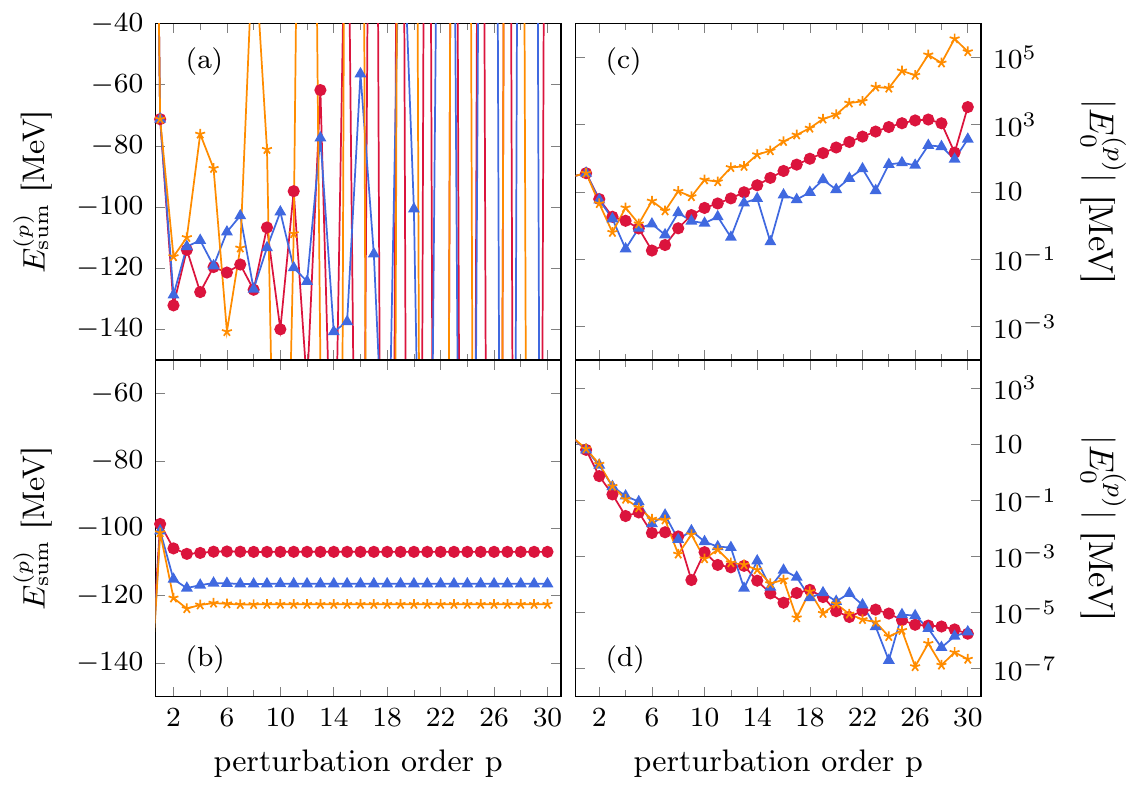}
\caption{Partial sums (left panels) and order-by-order corrections (right panels) for the ground-state energy of $^{16}$O using the Hamiltonian described in Sec.~\ref{sec:ham}. Top (bottom) panels correspond to the HO (HF) partitioning. All calculations are performed using $N_\text{max}=2,4,6$ (\redcircle,\bluetriangleup,\orangestar) and an oscillator frequency of $\hbar \omega=24\, \text{MeV}$. The SRG parameter is set to $\alpha=0.08\,\text{fm}^4$. Figure taken from Ref.~\cite{Tichai2016}.}
\label{fig:MBPT_hohf}
\end{figure}

\subsubsection{Impact of partitioning}

While perturbation theory defines a general framework to access nuclear observables, the performance strongly depends on the choice of the partitioning $H = H_0 + H_1$ or, equivalently, on the underlying vacuum fixing the starting point for the expansion. Subsequently, two choices for $H_0$ are presently compared in the calculation of the ground-state energy of $^{16}$O, i.e the one-body (i)  spherical harmonic oscillator (HO) and (ii) self-consistent HF\footnote{The HF problem is solved in a symmetry-restricted way, enforcing rotational invariance of the resulting single-particle basis.} Hmiltonians (cf. Sec~\ref{sec:partitioning}). The model space is truncated employing the $N_\text{max}$-truncation similar to the NCSM. Panel (a) of Fig.~\ref{fig:MBPT_hohf} shows the sequence of partials sums using a HO partitioning for a set of model spaces. The partial sums are divergent in all cases, which can equally be seen from the exponential divergence of high-order energy corrections in panel (c). On the other hand, using a HF reference state yields a rapidly converging perturbation series (panel (b)) and the energy corrections are exponentially suppressed as a function of the perturbative order (panel (d)), indicating robust convergence. In all cases the converged results agree up to numerical accuracy with the exact CI diagonalization.

Obviously, the reference state heavily affects the performance of MBPT. In the above case, this can be understood by the poor quality of the HO reference, e.g. the wrong asymptotic radial dependence of single-particle HO eigenstates (Gaussian instead of exponential suppression).  Consequently, in the following a HF determinant is used as a reference state when results are reported for closed-shell nuclei.

\subsubsection{SRG dependence \label{sec:srgresults}}

\begin{figure}[t!]
\centering
\includegraphics[width=0.75\textwidth]{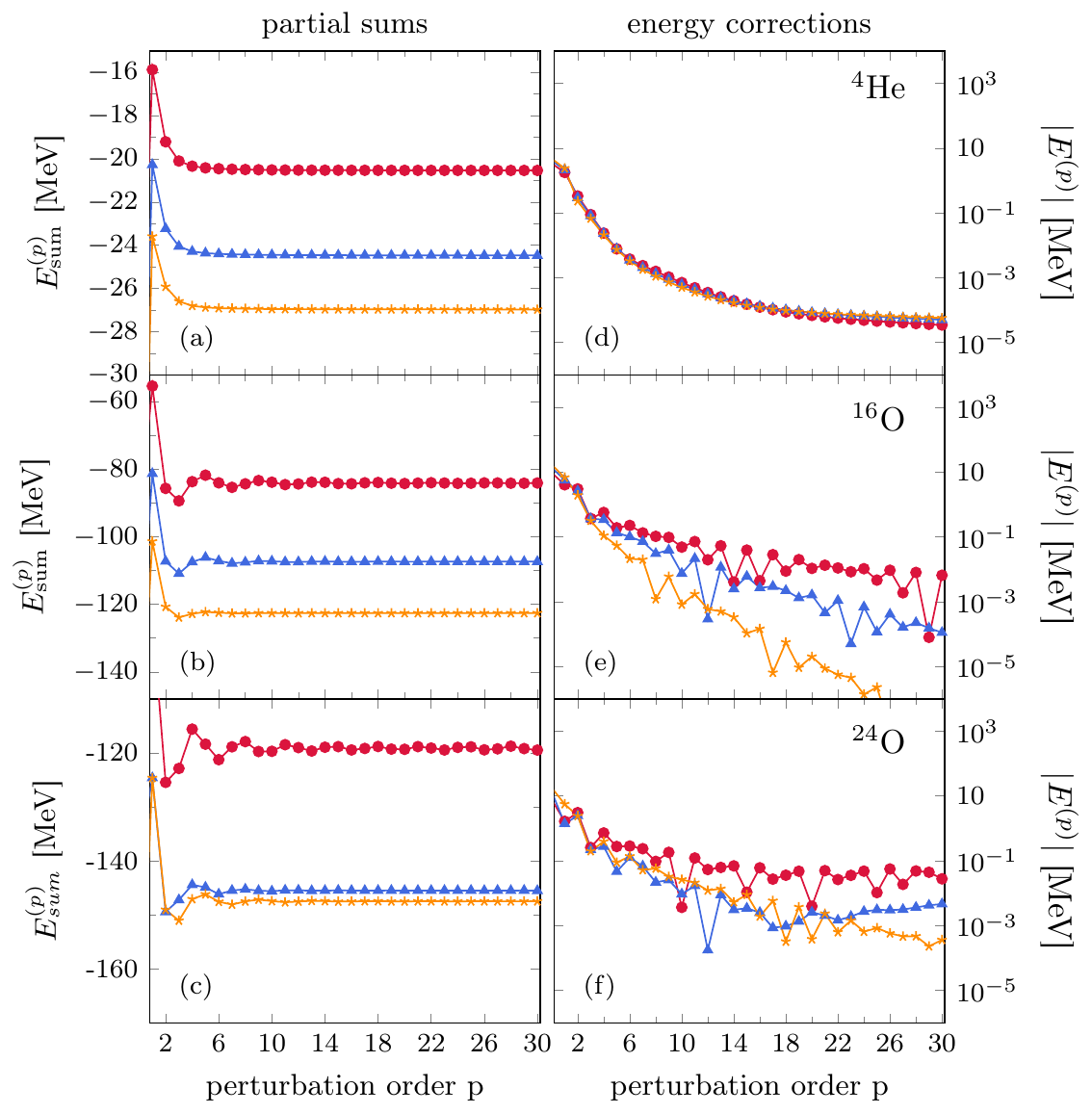}
\caption{Partial sums (left panels) and order-by-order corrections (right panels) corresponding to the ground-state energy obtained from the Hamiltonian described in Sec.~\ref{sec:ham}. Top, middle and bottom panels correspond to  $^4$He ($N_\text{max} = 6$), $^{16}$O ($N_\text{max} = 6$) and $^{24}$O ($N_\text{max} = 4$), respectively.
All calculations are performed setting the oscillator frequency to $\hbar \omega=20\, \text{MeV}$.  The different plot markers correspond to SRG flow parameter $\alpha = 0.02($\redcircle$), 0.0625$(\bluetriangleup) and $0.08\,\text{fm}^4$(\orangestar). 
Figure taken from Ref.~\cite{Tichai2016}.}
\label{fig:MBPT_srg}
\end{figure}

Using a HF partitioning, the impact of the SRG transformation of the Hamiltonian on the perturbative series is now illustrated. In Fig.~\ref{fig:MBPT_srg} the ground-state energy of $^4$He, $^{16}$O and $^{24}$O is displayed while varying the value of the flow parameter $\alpha$ defining the SRG transformation. The left-hand panels show that the perturbative series converge in all cases thus demonstrating the reliability of HF-MBPT. For the light $^4$He, the results are independent of the flow parameter and the MBPT expansion converges rapidly in all cases. For $^{16}$O and $^{24}$O, the rate of convergence is slower for harder interactions, i.e., for lower values of $\alpha$. Furthermore, the partial sums admit a damped oscillatory behaviour in the oxygen isotopes for $\alpha=0.02\, \text{fm}^{4}$. These features can be better seen from the right-hand panels, where lower values of $\alpha$ induce a slower suppression of higher-order corrections (panels (e) and (f)).

\subsubsection{Realistic calculations}

The previous results reveal that using an optimized HF reference state combined with a sufficiently soft interaction defines a well-controlled regime where perturbation theory can be robustly applied to closed-shell nuclei. For heavier nuclei and larger model spaces the high-order perturbative series cannot be computed recursively and, rather, low-order expressions are evaluated via explicit single-particle summations\footnote{Using the notion of tensors, the summation over a common index of several tensors defines a \emph{tensor contraction}.}. 

\begin{figure}[t!]
\centering
\includegraphics[width=0.9\textwidth]{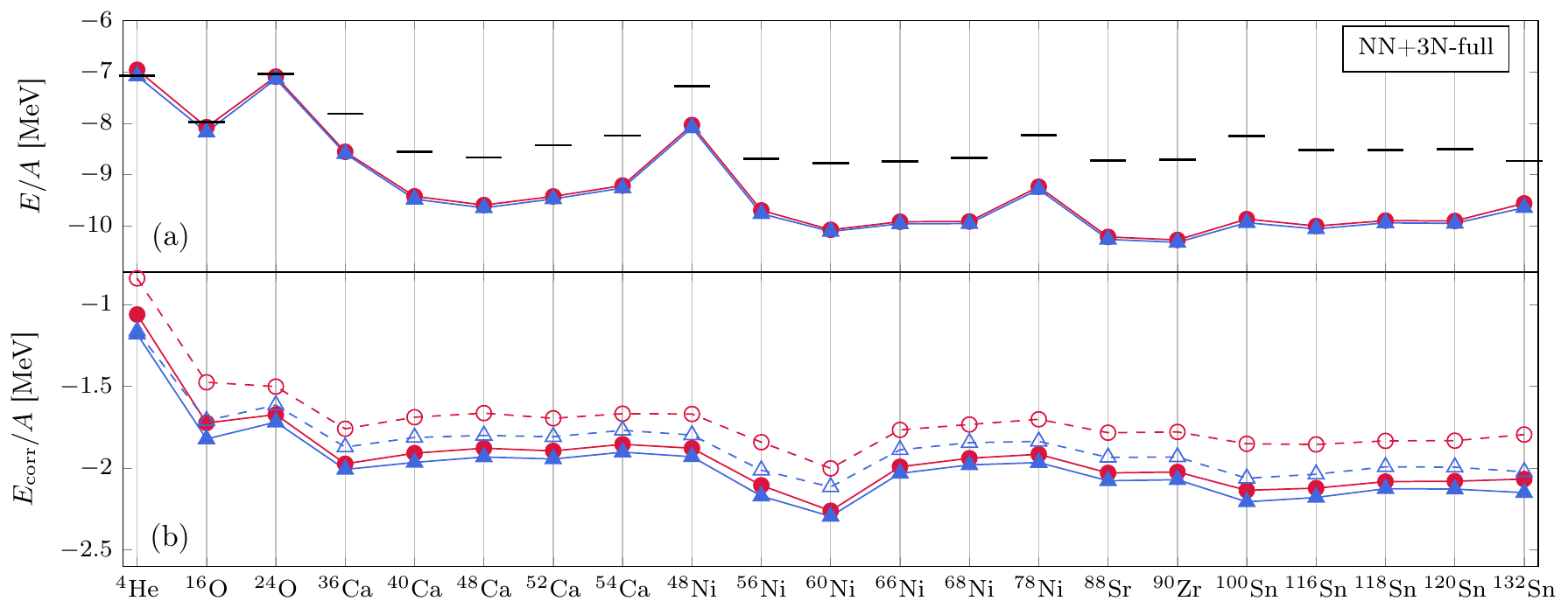}
\caption{Ground-state binding energies per particle (top panel) and correlation energy per particle (bottom panel) of closed-shell nuclei from second-order MBPT (\redcircleopen), third-order MBPT (\redcircle), CCSD (\bluetriangleupopen) and CR-CC(2,3) (\bluetriangleup) using the Hamiltonian described in Sec.~\ref{sec:ham}. 
All calculations are performed using 13 oscillator shells and an oscillator frequency of $\hbar \omega=20\, \text{MeV}$. The SRG parameter is set to $\alpha=0.08\,\text{fm}^4$. 
Figure taken from Ref.~\cite{Tichai2016}.}
\label{fig:MBPT_midmass_3nfull}
\end{figure}

In Fig.~\ref{fig:MBPT_midmass_3nfull}, the ground-state energy per particle (top panel) and the correlation energy per particle (bottom panel) of a selection of doubly closed-shell nuclei ranging from $^4$He to $^{132}$Sn is displayed at second- and third-order in MBPT~\cite{Tichai2016}. A model space built out of 13 major harmonic oscillator shells is employed. Since the target nuclei are out of reach of exact diagonalization, MBPT results are compared to state-of-the CC calculations employing the same input Hamiltonian and the same HF determinant reference state.

The top panel demonstrates that third-order calculations fully capture the bulk part of the ground-state energy and are in remarkable agreement with more sophisticated non-perturbative CC results, i.e. the deviation with CR-CC(2,3) is less than $1\%$ in all cases. A more refined analysis can be deduced from the bottom panel where the mean-field binding energy has been subtracted. The correlation energy accounts in most mid-mass systems for about $1-2\,\text{MeV}/A$ such that the reference HF results capture roughly $60-70\%$ of the overall binding energy. The CCSD correlation energy lies between second- and third-order results even though the CCSD wave function resums correlation effects beyond third order, thus indicating a repulsive effect on the binding from 2-particle/2-hole-excitations at $4$th order and beyond. When (approximately) including triple excitations through CR-CC(2,3), slightly stronger binding than in third-order MBPT is generated. 

Of course, the enormous benefit of low-order MBPT is that it excellently reproduces highly sophisticated CC results at a computational cost that is two orders of magnitude lower due to its non-iterative character.

\section{Multi-configurational perturbation theory}
\label{sec:mcpt}

Driven by the capacity of HF-MBPT to grasp dynamical correlations induced in closed-shell systems by soft chiral Hamiltonians, extensions to genuine open-shell systems were envisioned. As alluded to in Sec.~\ref{sec:div}, the presence of degeneracies with respect to particle-hole excitations does not allow the use of a single symmetry-conserving Slater-determinant reference state. A first possibility is to use a multi-determinantal reference state, which was shown to be very useful in electronic structure applications~\cite{Rolik2003,Surjan2004} and as a pre-processing tool in NCSM calculations~\cite{Roth2009}.

\subsection{Rationale}

The starting point of multi-configurational perturbation theory (MCPT) is the definition of an initial reference space 
\begin{align}
\mathcal{M}_{\text{ref}} \equiv \text{span} \big \{ | \Phi_\nu\ra \big \}
\end{align}
built from a set of orthonormal many-body Slater determinants $|\Phi_\nu\rangle$. The extended reference space redefines the nature of the $P$-space (and thus of the $Q$-space) introduced in Eq.~\eqref{space_partitioning}. The multi-configurational reference state $\psiref$ is chosen to be a normalized vector obtained from a diagonalization in $\mathcal{M}_\text{ref}$, i.e.,
\begin{align}
\psiref \equiv \sum_{\nu \in \mathcal{M}_\text{ref}} c_\nu \, \vert \Phi_\nu \rangle \, ,  \label{expref}
\end{align}
where $c_\nu$ denotes the expansion coefficients, typically obtained from a NCSM calculation. The initial diagonalization provides a set of \emph{non-degenerate} but multi-determinantal reference states carrying good symmetry quantum numbers. At the price of giving up the product-type character of the reference state, the degeneracy is lifted and a well-defined perturbative expansion can be designed~\cite{Tichai:2018ncsmpt}.

\remark{Convention}{The symbol $|\Psi\ra$ is used to emphasize the multi-determinantal character of the reference state and distinguish it from a single product state.  Furthermore, two different  notations are employed to designate the Slater determinants spanning the complete Hilbert space:
(i) Slater determinants belonging to $\mathcal{M}_\text{ref}$ are denoted by $|\Phi_\nu\ra$, i.e., as a capital Greek letter carrying a lower-case Greek index, whereas  
(ii) Slater determinants outside $\mathcal{M}_\text{ref}$ are denoted by $|\phi_i\ra$, i.e., as a lower-case Greek letter carrying a lower-case Roman index.}

It is worth noting that MCPT naturally accesses excited states by building the perturbation theory on top of the various vectors produced through the prior diagonalization in $\mathcal{M}_\text{ref}$. Of course, it is not guaranteed that the energetically lowest state in the initial NCSM calculation eventually corresponds to the ground state in the fully correlated limit, i.e.,  perturbative corrections may induce \emph{level crossings} among the various states.

\subsection{Partitioning}

Formally, the unperturbed Hamiltonian is written in the spectral representation as
\begin{align}
H_0 \equiv \sum_{k \in \mathcal{M}_\text{ref}} E^{(0)}_k | \Psi_k \ra \la \Psi_k \vert + \sum_{i \notin \mathcal{M}_\text{ref}} E^{(0)}_{i}\; \vert \phi_i \ra \la \phi_i \vert \, ,
\label{Hzero}
\end{align}
where $|\Psi_k\ra$ denote the NCSM eigenvectors within $\mathcal{M}_\text{ref}$, including  the particular one, e.g. the lowest state, playing the role of the reference state. Consequently, one obtains the following eigenvalue relation for the unperturbed Hamiltonian
\begin{align}
H_0 \,|\Psi_k\ra= E^{(0)}_k \, |\Psi_k\ra \, ,
\end{align}
since the set of determinants $\{ |\Phi _i\ra \notin \mathcal{M}_\text{ref} \}$ is orthogonal to $| \Psi_k\ra$. In principle, the construction of $H_0$ requires a \emph{full diagonalization} in the reference space, i.e., the solution of all eigenvectors and eigenvalues making the construction of the unperturbed solution rapidly unfeasible if the dimension of the reference state grows significantly. As will become clear below, the computation of the lowest-order correction only require to access the reference state, which is thus the only the eigenvector that needs to be solved for explicitly in this case. In that case one may resort to Lanczos algorithms, thus targeting a limited number of extremal eigenstates.

Zeroth-order energies $E^{(0)}_{i}$ of the unperturbed Slater determinants making up the $Q$ space, i.e. $\{ |\Phi _i\ra \notin \mathcal{M}_\text{ref} \}$, are given by the sum  of occupied single-particle energies defined as diagonal matrix elements of the one-body Hamiltonian
\begin{align}
h_{pq} \equiv t_{pq} + \sum_{rs} \ame{pr}{qs} \rho^{(0)}_{rs} \;, \label{baranger_h}
\end{align}
where the one-body density matrix of the reference state\footnote{If the density matrix involved in Eq.~\eqref{baranger_h} were the one of the fully correlated eigenstate of $H$, the one-body operator $h$ would be nothing but the \emph{Baranger Hamiltonian}~\cite{baranger70a,Duguet:2014tua}. Whenever the reference state reduces to the HF Slater determinant, $h$ identifies with the HF one-body Hamiltonian.} is introduced
\begin{align} 
\rho^{(0)}_{pq} = \psirefbra c^\dagger_p c_q   \psiref   \, .
\end{align}
In principle, an explicit three-body term can be included as well at the price of invoking the two-body density matrix. However, for the sake of computational simplicity a normal-ordered two-body (NO2B) approximation is employed  to approximately account for the inclusion of 3N interactions~\cite{Gebrerufael2016}. 

The zeroth-order energy of the reference state is also defined via the single-particle energies defined in Eq.~\eqref{baranger_h} while taking into account the multi-determinantal character of the reference state through the mean occupation of single-particle states, i.e., the diagonal elements of the one-body density matrix $\rho^{(0)}_{pp}$, so that
\begin{align} 
E_\text{ref}^{(0)} = \sum_p \epsilon_p \rho^{(0)}_{pp}  \, .
\end{align}

\subsection{Low orders}

With the partitioning of the Hamiltonian defined above, zeroth- and first-order MCPT contributions to the energy read as
\begin{align}
E^{(0)} &= \psirefbra H_0 \psiref  = E_\text{ref}^{(0)} \;, \\
E^{(1)} &= \psirefbra H_1 \psiref  = \psirefbra H \psiref - E_\text{ref}^{(0)} \;, 
\end{align}
such that their sum reproduces the full reference energy $E_\text{ref}^\text{NCSM}$ obtained via the diagonalization of the full Hamiltonian $H$ in $\mathcal{M}_{\text{ref}}$.
 
The second-order energy correction reads similarly to the one at play in standard MBPT, i.e. 
\begin{align}
E^{(2)} 
= -\sum_{i \notin \mathcal{M}_{\text{ref}}} \frac{\vert \psirefbra H \vert \phi_i\rangle \vert^2}{E^{(0)}_i - E_{\text{ref}}^{(0)} } \, ,
\end{align}
where the sum runs over states outside of the reference space and where the contribution from $H_0$ vanishes by orthogonality $\la \Psi_{\text{ref}} | \Phi_i \ra = 0$. To explicitly evaluate $E^{(2)}$ the reference state is expanded according to Eq.~\eqref{expref}
\begin{align}
E^{(2)}  = -\sum_{\mu \mu' \in \mathcal{M}_{\text{ref}}} c_\mu^\star c_{\mu^\prime} 
 \sum_{i \notin \mathcal{M}_{\text{ref}}} \frac{\langle \Phi_\mu  \vert H \vert \phi_i\rangle\langle \phi_i \vert H \vert \Phi_{\mu^\prime}\ra}{E^{(0)}_i - E_{\text{ref}}^{(0)}} \, .
\label{eq:mcpt2}
\end{align}

All many-body matrix elements appearing in the algebraic expressions of the perturbative corrections involve Slater determinants only and can be readily evaluated using standard NCSM technology. As an efficient alternative, normal-ordering techniques and standard Wick's theorem are employed such that an associated diagrammatic can be designed. It is worth noting that intermediate states from within $\mathcal{M}_{\text{ref}}$ only start contributing at fourth order~\cite{Tichai:2018ncsmpt} such that they do not appear in the evaluation of $E^{(2)}$. In Eq.~\eqref{eq:mcpt2}, the Hamiltonian is normal ordered with respect to the rightmost determinant $\vert \Phi_{\mu^\prime} \ra$ for each term in the sum over $\mu'$ and the two matrix elements are evaluated using the associated Wick's theorem. Similar techniques have been applied in quantum chemistry \cite{Hose1979,Hose1980}. The computational scaling of the second-order correction for large reference spaces is given by $\dim(\mathcal{M}_\text{ref})^2 \cdot n_p^2 \cdot n_h$, where $n_p$ and $n_h$ denote the number of particle and hole states, respectively.

\subsection{Results}

In the following the performance of MCPT, specifically denoted as NCSM-PT in the present case, is gauged in a similar spirit as for HF-MBPT in Sec.~\ref{sec:HFMBPT}.

\subsubsection{High-order corrections in light systems}
\label{highordersMCPT}

\begin{figure}[t!]
\centering
\includegraphics[width=0.75\textwidth]{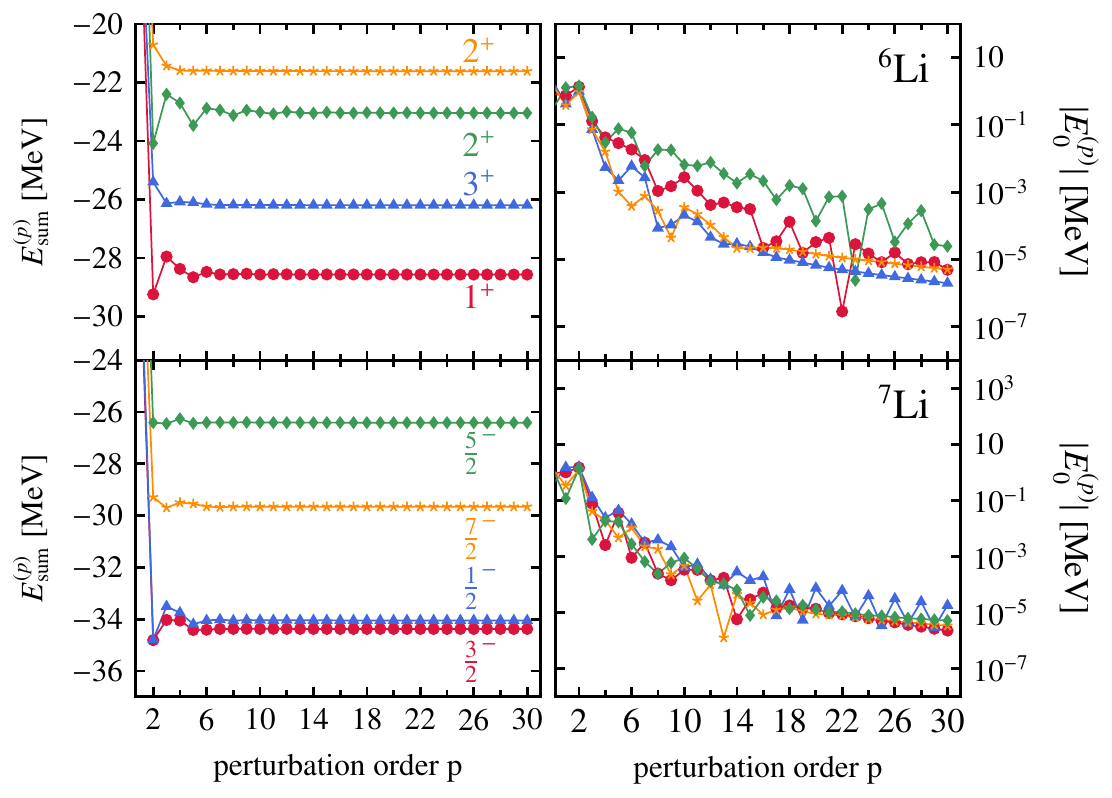}
\caption{High-order binding energies from NCSM-PT employing $\Nmaxref=0$ reference states using the Hamiltonian described in Sec.~\ref{sec:ham}. All calculations are performed including configurations up to $N_\text{max} = 8$ and employ an oscillator frequency of $\hbar \omega=20\, \text{MeV}$. The SRG parameter is set to $\alpha=0.08\,\text{fm}^4$. 
Figure taken from Ref.~\cite{Tichai:2018ncsmpt}.}
\label{fig:MCPT_highorder}
\end{figure}

The recursive treatment of HF-MBPT laid out in Sec.~\ref{sec:recursive} has proven invaluable to understand the convergence characteristic of the perturbative expansion. While NCSM-PT does not employ a Slater-determinant reference, the recursive formulation can be extended in a straightforward way~\cite{Tichai:2018ncsmpt}. Figure~\ref{fig:MCPT_highorder} displays the convergence behaviour of the perturbative series for $^{6,7}$Li built on top of the four lowest states of the $N_\text{max}=0$ NCSM diagonalization. The left-hand panels show that the perturbative series is convergent for both systems and all target states with a slight overbinding of the second-order partial sum for many states. In all cases the converged results agree with exact diagonalization. Furthermore, right panels reveal an (almost) exponential suppression of higher-order energy corrections indicating rapid convergence of the  expansion. The rate of convergence is mostly independent of the target state or the nucleus, except for the ground and  first $2^+$ states in $^6$Li that both converge slightly slower.

The high-order benchmarks strongly motivate the use of low-order partial sums as good approximations to the binding energies of heavier systems. Subsequently, systems with mass number $A\approx30$ are investigated through realistic MCPT calculations in large model spaces.

\subsubsection{Low orders in $A\approx30$ nuclei}

\begin{figure}[t!]
\centering
\includegraphics[width=1.0\textwidth]{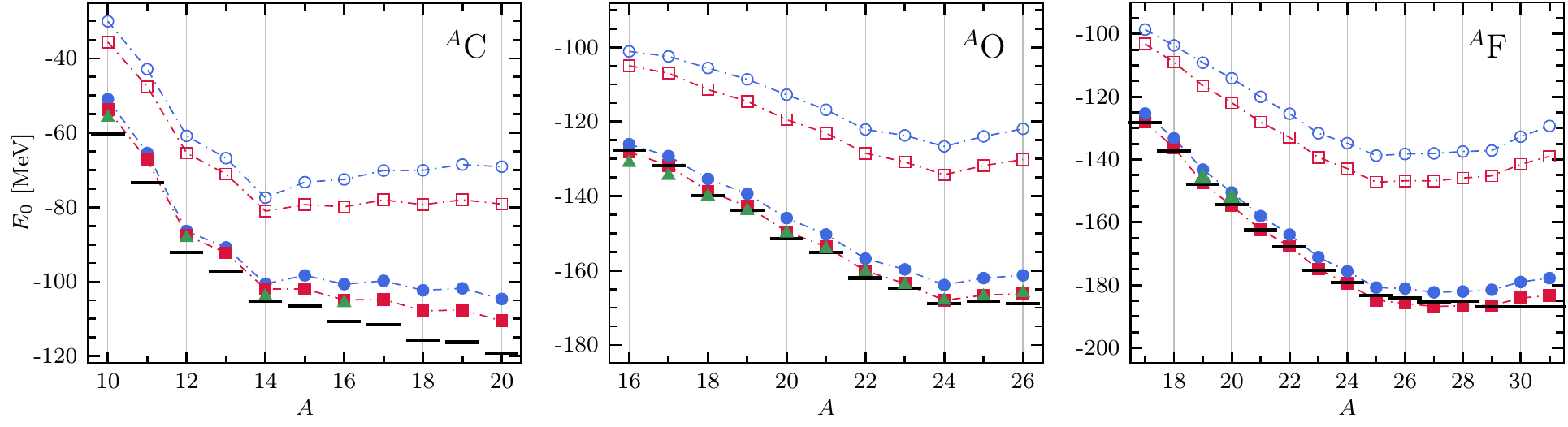}
\caption{Reference (\bluecircleopen  \slash \,\redsquareopen) and second-order NCSM-PT (\bluecircle  / \redsquare) energies with $\Nmaxref=0$ and $2$, respectively, for the ground states of $^{11-20}$C, $^{16-26}$O and $^{17-31}$F using the Hamiltonian described in Sec.~\ref{sec:ham}. 
All calculations are performed using 13 oscillator shells and an oscillator frequency of $\hbar \omega=20\, \text{MeV}$.
The SRG parameter is set to $\alpha=0.08\,\text{fm}^4$. 
Importance-truncated NCSM calculations (\greentriangleup) are shown for comparison.
Experimental values are indicated by black bars.
Figure taken from Ref.~\cite{Tichai:2018ncsmpt}.}
\label{fig:MCPT_CFO}
\end{figure}

In Fig.~\ref{fig:MCPT_CFO} ground-state energies of carbon, oxygen and fluorine isotopes are shown and compared to large-scale importance truncated NCSM (IT-NCSM) diagonalization whenever available. The reference states are obtained from a diagonalization in a $N_\text{max}=0$ or $2$ space. In all cases a HF single-particle basis is employed in order to minimize the dependence on the oscillator frequency\footnote{The HF problem is solved employing a so-called \emph{equal-filling approximation} to ensure rotational invariance of the mean-field density. The underlying NCSM-PT formulation is based on $m$-scheme quantities and does not employ angular-momentum coupling techniques as most correlation expansions do. Consequently, even- and odd-mass systems can be described on equal footing.}.

For all nuclei the reference energies and second-order NCSM-PT results show a sizeable dependence on the size of the reference space. Throughout all investigated isotopic chains though, results from reference states built within a $\Nmaxref=2$ space almost perfectly reproduce the exact (IT-)NCSM ones. This significant improvement in ground-state energies hints at important correlations incorporated through the reference states obtained from a $\Nmaxref=2$ diagonalization that are absent for $\Nmaxref=0$, thus providing an ideal compromise between computational efficiency and accuracy. In particular, neutron-rich fluorine isotopes are out of reach of conventional NCSM calculations and NCSM-PT provides an efficient \emph{ab initio} approach to investigate the neutron drip line. A single NCSM-PT calculation requires typically two to three orders of magnitude less computational resources than the corresponding IT-NCSM calculation.

In practice, the reference states employed in the above calculations contain between several hundreds of thousands up to a few million determinants, thus, providing an excellent account of static correlations.
Note that in most cases the reference state accounts for up to $80\%$ of the overall binding energy such that residual dynamical correlations can indeed be grasped efficiently from low-order perturbation theory.

\subsubsection{Low-lying spectroscopy}

\begin{figure}[t!]
\centering
\includegraphics[width=0.95\textwidth]{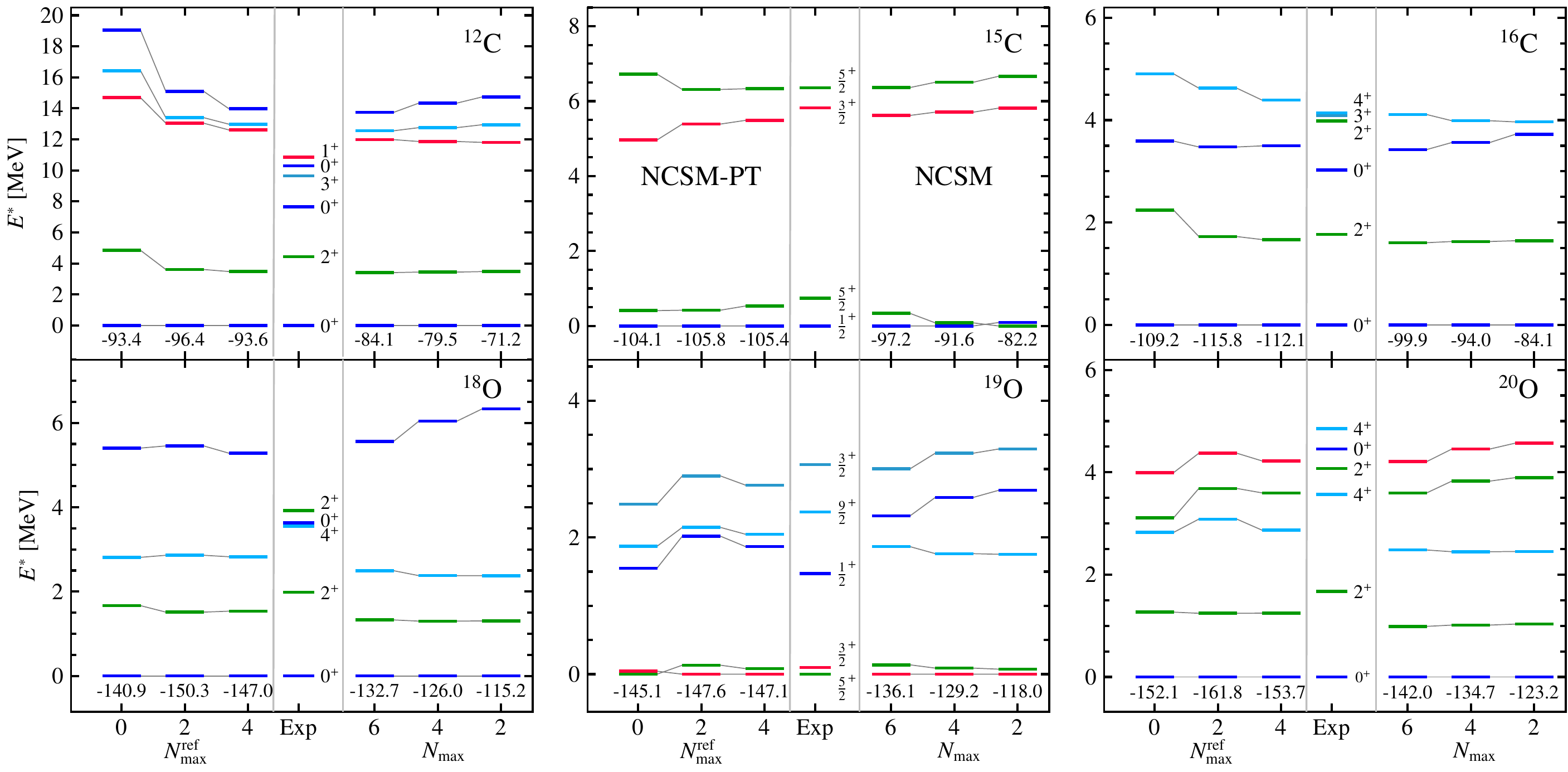}
\caption{Spectra obtained via second-order NCSM-PT for selected carbon and oxygen isotopes using the Hamiltonian described in Sec.~\ref{sec:ham}.  All calculations are performed using 13 oscillator shells and an oscillator frequency of $\hbar \omega=20\, \text{MeV}$. The SRG parameter is set to $\alpha=0.08\,\text{fm}^4$. A HO single-particle basis was used to separate center-of-mass contaminations. Large-scale IT-NCSM calculations are shown for comparison. Figure taken from Ref.~\cite{Tichai:2018ncsmpt}.}
\label{fig:MCPT_spectra}
\end{figure}

As already exemplified in the high-order investigation, excited states can be straightforwardly accessed through NCSM-PT by targeting different reference states from the NCSM spectrum. From absolute NCSM-PT binding energies, excitation energies are obtained by subtracting the correlated ground-state energy. In Fig.~\ref{fig:MCPT_spectra} the associated NCSM-PT spectra are compared to bare NCSM calculations for a selection of open-shell carbon and oxygen isotopes. All calculations employ a HO single-particle basis to separate center-of-mass degrees of freedom in the many-body wave function. It is well-known that NCSM excitation energies of states with identical parity display a much faster convergence than absolute binding energies, thus yielding stable results in the right-hand columns of each panel. When a level re-ordering appears with increasing $N_\text{max}$ in the NCSM calculation, e.g. for the two lowest states in $^{12}$C or the third and fourth states in $^{19}$O, NCSM-PT reproduces the correct level ordering at small values of $\Nmaxref$. As for ground-state energies, NCSM-PT results based on a reference state with $\Nmaxref=2$ reproduce well the NCSM spectra, while going to $\Nmaxref=4$ only refines the quality of a subset of excitation energies.

The NCSM-PT framework is thus highly valuable to perform light-weighted perturbative calculations in medium-light systems up to mass numbers $A \lesssim 40$. Due to its versatility and conceptual simplicity the low-lying spectrum of genuine open-shell nuclei can be described at low computational cost.

\section{Bogoliubov many-body perturbation theory}
\label{sec:BMBPT}

While the use of multi-configurational reference states can efficiently resolve situations of strong static correlations, it displays several limitations, i.e. (i) the physical origin of the underlying correlations in the reference state is unclear, (ii) it does not easily ensure size-extensivity and (iii) it is numerically prohibitive in heavy nuclei. The objective is thus to present an alternative based on single-reference product states that bypasses these limitations.

\subsection{Rationale}

An alternative route to lift the particle-hole degeneracy of the reference state in open-shell systems is to authorize the reference state to break a symmetry of the underlying Hamiltonian. In semi-magic nuclei, the relevant symmetry is $U(1)$ global gauge symmetry associated with particle-number conservation\footnote{In fact the relevant symmetry group is the direct product $U_N(1)\times U_Z(1)$ since both proton and neutron number are conserved separately.}. Breaking $U(1)$ symmetry permits to efficiently deal with Cooper pair instability associated with the superfluid character of open-shell nuclei.  The degeneracy of a Slater determinant with respect to particle-hole excitations is lifted via the use of a Bogoliubov reference state and transferred into a degeneracy with respect to transformations of the symmetry group. As a consequence, the ill-defined (i.e.\@ singular) expansion of exact quantities with respect to a symmetry-conserving Slater determinant is replaced by a well-behaved one. Extending the treatment to doubly open-shell nuclei requires a similar treatment of $SU(2)$ symmetry associated with the conservation of angular momentum.

Eventually, the degeneracy with respect to $U(1)$ transformations must also be lifted by restoring the symmetry.  However, BMBPT only restores the symmetry in the limit of an all-order resummation, and thus displays a symmetry contamination at any finite order. While BMBPT can still be used as a stand-alone approach as is done in the present work, it eventually provides the first step towards the implementation of the so-called particle-number projected BMBPT (PNP-BMBPT)~\cite{Duguet:2015yle} that restores good particle number at any truncation order.

\subsection{Bogoliubov algebra}

The BMBPT formalism is based on the introduction of the Bogoliubov reference state
\begin{align}
\HFB \equiv \mathcal{C} \prod_k \beta_k \vert 0 \ra\, ,
\end{align}
where $\mathcal{C}$ is a complex normalization constant and $\vert 0 \ra$ denotes the physical vacuum. The Bogoliubov product state presently defining the $P$-space of perturbation theory is a vacuum for the quasi-particle operators $\{\beta^\dagger_k, \beta_k\}$, i.e.,
\begin{align}
\beta_k | \Phi \ra =0 \, \quad \forall k \, ,
\end{align}
obtained from the creation and annihilation operators associated with a basis of the one-body Hilbert space via the unitary Bogoliubov transformation~\cite{RiSc80}
\begin{subequations}
\begin{align}
\beta_k &\equiv \sum_p U^*_{pk} c_p + V^*_{pk} c^\dagger_p\, , \\
\beta_k^\dagger &\equiv \sum_p U_{pk} c^\dagger_p + V_{pk} c_p \, .
\end{align}
\end{subequations}

Generically speaking, the Bogoliubov transformation is only constrained by unitarity such that a large manifold of Bogoliubov states is at hand. To actually set up the perturbation theory, a particular Bogoliubov reference state must be specified. Typically, the transformation matrices $(U,V)$ are obtained by solving the Hartree-Fock-Bogoliubov (HFB) variational problem that naturally extends the simpler HF approximation to treat pairing correlations while sticking to a single reference product state. The columns of the transformation matrices $(U,V)$ correspond to the eigenvectors of the HFB eigenvalue equation~\cite{RiSc80} whereas the associated eigenvalues $\{E_k > 0\}$ deliver the so-called quasi-particle energies\footnote{In the present work, the Bogoliubov transformation is limited to treat the like-particle pairing although it could be further generalized to address neutron-proton pairing as well.}. Since Bogoliubov states are not eigenstates of the particle-number operator $A$, the expectation value of $A$ is constrained to match a specific number of particles, e.g. the particle number $A_0$ of the target system. In the HFB method for example, the constraint is enforced via the use of a Lagrange multiplier $\lambda$ in the minimization of the expectation value of the grand potential 
\begin{align}
\Omega \equiv H - \lambda A \, .
\end{align}
In actual applications, separate Lagrange multipliers $\lambda_N$ and $\lambda_Z$ are used to constrain proton and neutron numbers $N$ and $Z$, respectively. In the subsequent formalism $A$ stands for either one of them.

\subsection{Quasi-particle normal ordering}

In the next step, Wick's theorem is employed to normal order the grand potential $\Omega$ with respect to the Bogoliubov reference state
\begin{align}
\Omega &= \underbrace{\Omega^{00}}_{\displaystyle  \equiv\Omega^{[0]}}  + 
 \underbrace{\Omega^{20} + \Omega^{11} +\Omega^{02}}_{\displaystyle\equiv  \Omega^{[2]}}  +   \underbrace{\Omega^{40} + \Omega^{31} +\Omega^{22} +\Omega^{13} +\Omega^{04}}_{\displaystyle  \equiv\Omega^{[4]}} + \ldots \, ,
\label{eq:NO}
\end{align} 
where $\Omega^{ij}$ denotes the normal-ordered component involving $i$ ($j$) quasi-particle creation (annihilation) operators, e.g., 
\begin{align}
\Omega^{31} &\equiv \frac{1}{3!}\sum_{k_1 k_2 k_3 k_4}  \Omega^{31}_{k_1 k_2 k_3 k_4}
   \beta^{\dagger}_{k_1}\beta^{\dagger}_{k_2}\beta^{\dagger}_{k_3}\beta_{k_4} \, .
\end{align} 
The tensors defining each normal-ordered term display antisymmetry properties, i.e.
\begin{align}
\Omega^{ij}_{k_1 \ldots k_{i} k_{i+1} \ldots k_{i+j}} = (-1)^{\sigma(P)}
\Omega^{ij}_{P(k_1 \ldots k_i | k_{i+1} \ldots k_{i+j})}  \, ,
\end{align}
where $\sigma(P)$ refers to the signature of the  permutation $P$.  The notation $P(\ldots | \ldots)$ denotes a  separation into the $i$ quasiparticle creation operators and the $j$ quasiparticle annihilation operators such that permutations are only considered among members of the same group.
Thus, $\Omega^{00}$ is the expectation value of $\Omega$ in $\HFB$ whereas $\Omega^{[2]}$ and $\Omega^{[4]}$ define effective, i.e., normal-ordered, one-body and two-body operators, respectively.
Working in the particle-number-conserving normal-ordered two-body approximation (PNO2B)~\cite{Ripoche2019}, the effective three-body part $\Omega^{[6]}$ is presently discarded\footnote{The particle-number-conserving nature of the PNO2B approximation further requires to drop specific contributions to  $\Omega^{[4]}$ as well; see Ref.~\cite{Ripoche2019} for a detailed discussion.}. Details on the normal-ordering procedure as well as expressions of the matrix elements of each operator $\Omega^{ij}$ in terms of the original matrix elements of the Hamiltonian and of the $(U,V)$ matrices can be found in Refs.~\cite{Si15,Ripoche2019}.

\subsection{Partitioning}

To set up the perturbation theory, the Hamiltonian (i.e.\@ grand potential) must be partitioned into an one-body unperturbed part $\Omega_{0}$ and a residual part $\Omega_{1}$, i.e.,
\begin{equation}
\label{split1}
\Omega = \Omega_{0} + \Omega_{1} \, .
\end{equation} 
Focusing on the case where the Bogoliubov reference state is the solution of the HFB variational problem, i.e. using a M{\o}ller-Plesset scheme, $\Omega$ appearing in Eq.~\eqref{eq:NO} is naturally partitioned given that 
\begin{align}
\Omega^{20}=\Omega^{02}=0
\end{align}
and that $\Omega^{11}$ is in diagonal form, i.e.,  
\begin{subequations}
\label{perturbation}
\begin{align}
\Omega_{0} &\equiv \Omega^{00} + \sum_{k} E_k \, \beta^{\dagger}_k \beta_k \label{perturbation1} \, , \\
\Omega_{1} &\equiv  \Omega^{40} + \Omega^{31} +\Omega^{22} +\Omega^{13} +\Omega^{04} \label{perturbation2} \, ,
\end{align}
\end{subequations}
with $E_k > 0$ for all $k$. Introducing all many-body states obtained via an even number of quasi-particle excitations of the vacuum
\begin{equation}
| \Phi^{k_1 k_2\ldots} \rangle \equiv \beta^{\dagger}_{k_1} \, \beta^{\dagger}_{k_2} \,  \ldots  |  \Phi \rangle \, , 
\end{equation}
the unperturbed system is fully characterized by its complete set of orthonormal eigenstates in Fock space
\begin{subequations}
\begin{align}
\Omega_{0}\, |  \Phi \rangle &= \Omega^{00} \, |  \Phi \rangle \, , \\
\Omega_{0}\, |  \Phi^{k_1 k_2\ldots} \rangle &= \left[\Omega^{00} + E_{k_1 k_2 \ldots}\right] |  \Phi^{k_1 k_2\ldots} \rangle  \label{phi} \, ,
\end{align}
\end{subequations}
where the strict positivity of unperturbed excitations 
\begin{align}
E_{k_1 k_2 \ldots} \equiv E_{k_1} + E_{k_2} +\ldots  
\end{align}
characterizes the lifting of the particle-hole degeneracy authorized by the spontaneous breaking of $U(1)$ symmetry in open-shell nuclei at the mean-field level.

With these ingredients at hand, the perturbation theory can be entirely worked out algebraically or diagrammatically.  This can be done on the basis of a (imaginary) time-dependent formalism or of a time-independent formalism.  While the former framework leads to working with Feynman (time-dependent) diagrams, the latter makes use of Goldstone (time-ordered) diagrams. Recently, the complete Rayleigh-Schr\"odinger BMBPT formalism, including the automatic generation and algebraic evaluation of all possible diagrams appearing at an arbitrary order $n$ on the basis of 2N and full 3N interactions has been published in Ref.~\cite{Arthuis:2018yoo}.

Eventually, the BMBPT expansion of the correlation energy can be written in compact form as a \emph{superfluid extension} of the Goldstone formula (Eq.~\eqref{eq:goldstoneE})
\begin{align}
\Delta \Omega = \la \Phi \vert \Omega \sum_{k=1}^\infty \Big( \frac{1}{\Omega^{00}- \Omega_1} \Omega_1 \Big)^{k} \HFB_c \, ,
\label{eq:Goldstone}
\end{align}
where the Hamiltonian is replaced by the grand potential.

\subsection{Low orders}

As a result of Wick's theorem with respect to $\HFB$, the first few orders contribute to Eq.~\eqref{eq:Goldstone}, with $\Omega^{(p)} \equiv E^{(p)} - \lambda A^{(p)}$ and $\Omega_\text{ref} = \Omega^{(0)}+\Omega^{(1)}$, according to\footnote{Non-canonical contributions are set to zero here given that we use the HFB reference state. When using an arbitrary Bogoliubov reference state, additional non-canonical contributions arise~\cite{Arthuis:2018yoo}.}
\begin{subequations}
\label{BMBPTcorrections}
\begin{align}
\Omega_\text{ref} &=  +\Omega^{00} \, ,  \\
\Omega^{(2)} &= -\frac{1}{24} \sum_{k_1 k_2 k_3 k_4} \frac{\Omega^{40}_{k_1k_2k_3k_4} \Omega^{04}_{k_3 k_4 k_1k_2} }{E_{k_1k_2k_3k_4}}  \, , \\
\Omega^{(3)} &= +\frac{1}{8} \sum_{k_1 k_2 k_3  k_4 k_5 k_6 } \frac{\Omega^{40}_{k_1k_2k_3k_4} \Omega^{22}_{k_3 k_4 k_5k_6 } \Omega^{04}_{k_5 k_6 k_1 k_2} }{E_{k_1 k_2 k_3 k_4} E_{k_5 k_6 k_1 k_2}}  \, . \label{BMBPTcorrections3}
\end{align}
\end{subequations}
The lifting of the degeneracy with respect to particle-hole excitations is embodied in the fact that the energy denominators in Eq.~\eqref{BMBPTcorrections} are non-singular and well behaved.  Indeed, quasi-particle energies are bound from below by the superfluid \emph{pairing gap} at the Fermi energy, i.e., 
\begin{align}
\text{Min}_{k} \{E_k\} \geq \Delta_{\text{F}} > 0 \, .
\end{align}
This would not be true in standard MBPT based on a Slater determinant reference state, where energy denominators associated with particle-hole excitations within the open shell would be zero in Eq.~\eqref{eq:mbpt2}.  Of course, BMBPT does strictly reduce to standard MBPT in a closed-shell system~\cite{Arthuis:2018yoo}.  In particular, the single third-order diagram whose algebraic expression is given in Eq.~\eqref{BMBPTcorrections3} generates the three, i.e., particle-particle, hole-hole and particle-hole, third-order HF-MBPT diagrams~\cite{Arthuis:2018yoo}. This reduction of the number of diagrams at any order $p$ is a consequence of working in a quasi-particle representation that does not distinguish particle and hole states. Conversely, all summations over quasi-particle labels run over the entire dimension of the one-body Hilbert space, which significantly increases the computational cost compared to standard MBPT. In any case, low-order BMBPT corrections only induce low polynomial scaling with respect to quasi-particle summation and do not suffer from the storage of large tensors as in more sophisticated all-order many-body approaches such as (B)CC or IMSRG. 

Extracting the $p$-order contribution to the binding energy from Eq.~\eqref{eq:Goldstone} requires the subtraction of the Lagrange term $\lambda A^{(p)}$. Computing $A^{(p)}$ can be done straightforwardly by replacing the leftmost operator $\Omega$ by $A$ in Eq.~\eqref{eq:Goldstone}~\cite{Arthuis:2018yoo}. As the reference state is constrained to have the correct particle number on average, it implies that $A_0^{(1) } = \text{A}_0$. Working with the HFB reference state, it can be shown that $A_0^{(2)}=0$ due to the fact that $\Omega^{20}=\Omega^{02}=0$. Consequently, the first correction to the average particle number appears at third order such that 
\begin{align}
A^{(1)}_0 + A_0^{(3)} \neq \text{A}_0 \, ,
\end{align}
i.e., the computed average particle number does not match the targeted number $A_0$ of the physical system. This feature requires an iterative BMBPT scheme in order for the particle number to be correct at the working order, e.g., $p \geq 3$, of interest~\cite{Demol:2020}.  To do so, one needs to rerun the HFB calculation with a $p$-dependent chemical potential such that, through a series of iterations, one eventually obtains, e.g., $A_0^{(1)} + \ldots + A_0^{(p)} = \text{A}_0$. Such a costly algorithm can fortunately be very well approximated by an \emph{a posteriori} correction scheme that entirely bypasses the iterative scheme~\cite{Demol:2020}. The third-order results presented in Sec.~\ref{loworderBMBPTresults1} have been computed without any adjustment of the average particle number whereas the novel ones discussed in Sec.~\ref{loworderBMBPTresults2} have been obtained on the basis of the \emph{a posteriori} correction scheme.

While the BMBPT expansion efficiently grasps static correlation effects associated to nuclear superfluidity, the breaking of a continuous symmetry in a finite quantum system is always fictitious.
Consequently, a full-fledged many-body formalism requires the additional restoration of the broken symmetry, i.e., a mixing of gauge-rotated Bogoliubov vacua which are connected to each other via (highly non-perturbative) symmetry transformations. While the formalism has already been laid out ~\cite{Duguet:2015yle}, realistic calculations remain yet to be performed. Still, proof-of-principle applications to the model pairing Hamiltonian employing particle-number-projected BCC theory~\cite{Qiu:2018edx} revealed the significant impact of the symmetry projection in the weakly broken regime corresponding to open-shell nuclei in the vicinity of shell closures.

\subsection{Results}

\subsubsection{Low-order calculations in mid-mass nuclei}
\label{loworderBMBPTresults1}

\begin{figure}[t!]
\centering
\includegraphics[width=0.9\textwidth]{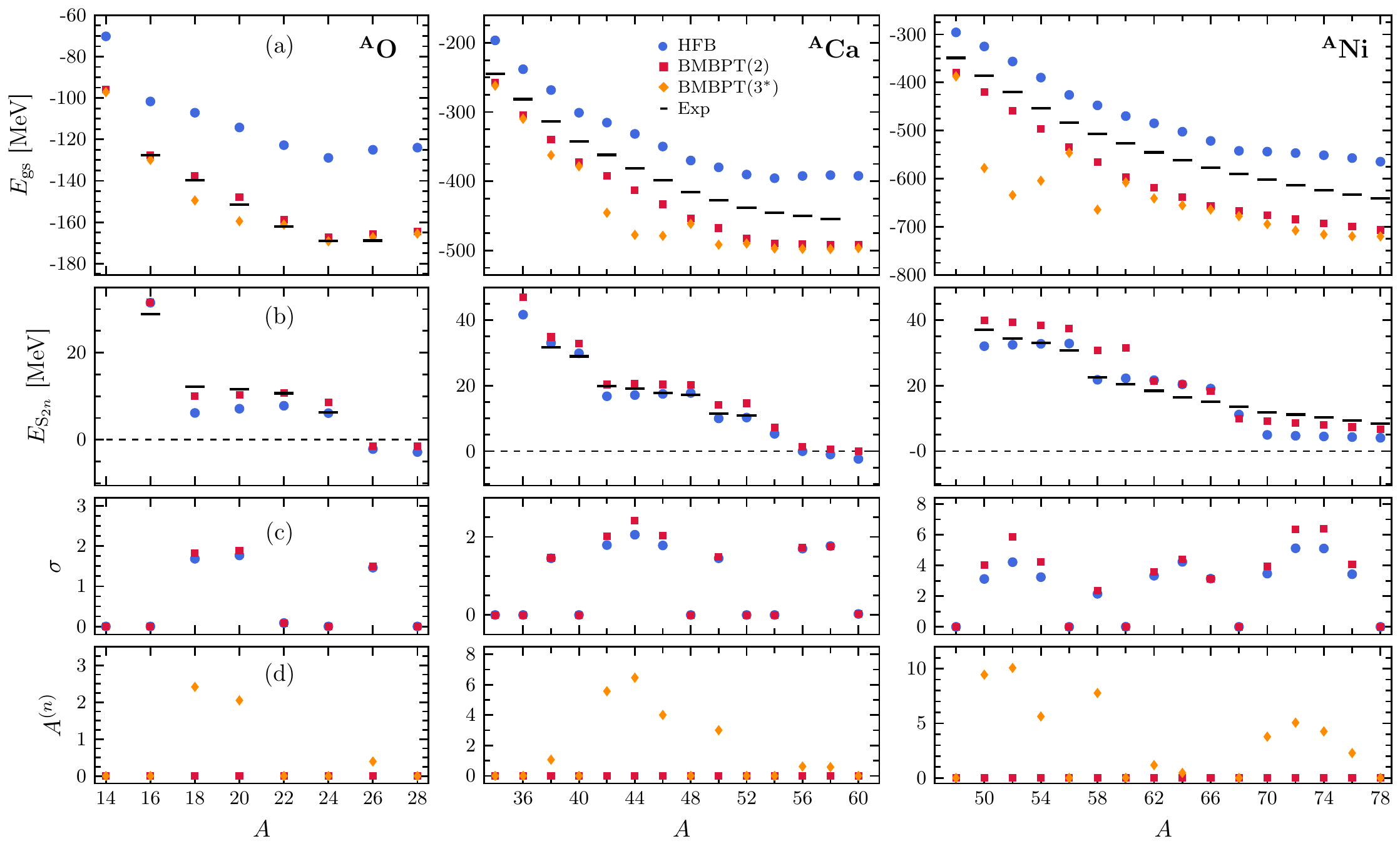}
\caption{BMBPT systematics along O, Ca and Ni isotopic chains: (a) absolute binding energies, (b) two-neutron separation energies, (c) neutron-number dispersion and (d) perturbative correction to the average neutron number. The employed Hamiltonian is defined in Sec.~\ref{sec:ham}.  Plot markers correspond to HFB (\bluecircle), second-order BMBPT (\redsquare) and third-order BMBPT without particle-number adjustment (\orangediamond). All calculations are performed using 13 oscillator shells and an oscillator frequency of $\hbar \omega=20\, \text{MeV}$. The SRG parameter is set to $\alpha=0.08\,\text{fm}^4$. Picture taken from 
Ref.~\cite{Tichai:2018mll}.}
\label{fig:BMBPT_detail}
\end{figure}

Figure~\ref{fig:BMBPT_detail} displays ground-state energies, two-neutron separation energies, particle-number variance and perturbative particle-number corrections along oxygen, calcium and nickel isotopic chains at the mean-field (i.e. HFB) level, as well as at second- and third-order in the BMBPT expansion. Top panels reveal that, while static correlations have partially been accounted for by employing a symmetry-broken reference state, the bulk of dynamical correlations are efficiently grasped via low-order BMBPT corrections. For (sub-)closed-shell nuclei the third-order correction is consistently suppressed indicating rapid convergence. In open-shell systems third-order partial sums are strongly contaminated due to a significant excess of neutrons brought by the third-order contribution to the average neutron number. As discussed earlier, calculations at third-order and beyond must eventually be done while constraining the average particle number to match the physical value~\cite{Demol:2020}. This is reported on in  Sec.~\ref{loworderBMBPTresults2}.

Panel (b) exhibits a qualitative reproduction of two-neutron separation energies already at the mean-field level. Results are quantitatively improved once second-order effects are incorporated. Panel (c) shows the neutron-number dispersion $\sigma \equiv \sqrt{ \la A^2 \ra - \la A\ra ^2}$ that grows with mass number. While the second-order contribution does not decrease yet the  neutron-number dispersion, one expects higher orders to do so~\cite{Demol:2020}. In closed-shell systems, the particle-number dispersion is zero as a hallmark of the particle-number-conserving character of the  wave function throughout the expansion.

A detailed study of the convergence characteristics of the BMBPT expansion via the calculation of high-order corrections similar to the ones presented in Sec.~\ref{sec:HFMBPT} (Sec.\ref{highordersMCPT}) for HF-MBPT (NCSM-PT) calculations of closed-shell (open-shell) nuclei has just be completed~\cite{Demol:2020} and is thus not reported on here.

\subsubsection{Comparison to non-perturbative calculations}

\begin{figure}[t!]
\centering
\includegraphics[width=0.9\textwidth]{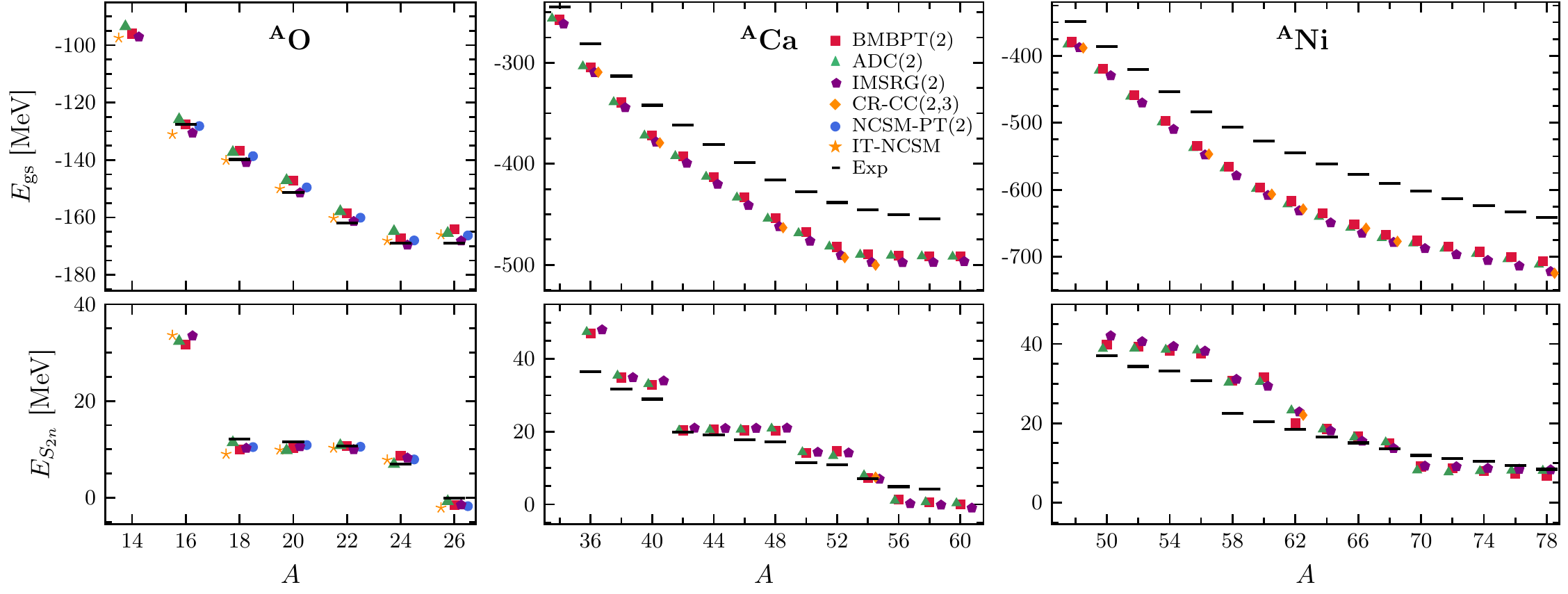}
\caption{Absolute ground-state binding energies (top) and two-neutron separation energies (bottom) along O, Ca and Ni isotopic chains using the Hamiltonian defined in Sec.~\ref{sec:ham}. 
Plot markers correspond to second-order BMBPT (\redsquare), second-order NCSM-PT (\bluecircle), large-scale IT-NCSM (\orangestar), GSCGF-ADC(2) (\greentriangleup), MR-IMSRG(2) (\hspace{1.7pt}\purplepentagon) and CR-CC(2,3) (\orangediamond).
All calculations are performed using 13 oscillator shells and an oscillator frequency of $\hbar \omega=20\, \text{MeV}$. The SRG parameter is set to $\alpha=0.08\,\text{fm}^4$.
Picture taken from Ref.~\cite{Tichai:2018mll}.}
\label{fig:BMBPT_comparison}
\end{figure}

Figure~\ref{fig:BMBPT_comparison} compares second-order BMBPT ground-state and two-neutron separation energies with results obtained from state-of-the-art non-perturbative many-body frameworks along oxygen, calcium and nickel isotopic chains. In particular, IT-NCSM provides essentially exact reference results along the oxygen chain. In heavier closed-shell nuclei, advanced CR-CC(2,3) calculations also provide reference results.

In all cases, second-order BMBPT is in excellent agreement with other methods, displaying deviations of less than $2\%$.  In particular, IT-NCSM results are very well reproduced in oxygen isotopes. While NCSM-PT (see Sec.~\ref{sec:mcpt}) and MR-IMSRG systematically generate stronger binding, GSCGF at the ADC(2) level is very close for all systems investigated. In the case of closed-shell nuclei the stronger binding obtained in the CR-CC(2,3) calculation highlights the importance of an (approximate) incorporation of 3-particle/3-hole excitations.

As seen from two-neutron separation energies, all \emph{ab initio} methods consistently predict the same shell structure and location of the neutron trip line. This feature highlights both the great success of recently developed many-body methods and the excellent performance of perturbative techniques such as NCSM-PT and BMBPT in particular. On the other hand, the strong deviation of absolute binding energies\footnote{The strong deviation also concerns charge radii that are not presently discussed.}  from experimental data, most pronounced in neutron-rich calcium and nickel isotopes, is common to all employed frameworks and clearly points to defects of the employed Hamiltonian. A detailed discussion of this crucial issue is postponed to Sec.~\ref{sec:newham}, where a new family of chiral Hamiltonians is tested with the goal to cure the poor agreement presently obtained for nuclear ground-state observables of mid-mass nuclei.

\section{Importance truncation}
\label{sec:it}

In previous sections, MBPT was considered as a full-fledged standalone framework to access the solution of the quantum many-body problem. However, MBPT techniques can also be used to support non-perturbative approaches by pre-processing the many-body configuration space. Subsequently, the concept of \emph{importance truncation} (IT) is presented as a procedure to a priori remove a set of tensor entries to be solved for on the basis of  an importance measure~\cite{RoNa07,Roth2009,Stumpf2016,Tichai:2019ksh}. Historically, this idea was first applied in electronic structure theory to perform a pre-selection of multi-reference CI amplitudes~\cite{Buenker1974,Buenker1975}.

\subsection{Original context: configuration interaction methods}
\label{ITNCSM}

The IT concept is ideally suited for basis expansion methods, such as general configuration interaction (CI) approaches or, more specifically, the NCSM. Given the expansion of an eigenstate $\ket{\Psi_n}$ of $H$ in some many-body basis $\ket{\phi_{i}}$, the importance of each basis state can be quantified through the associated expansion coefficient $C_{i}^{(n)} \equiv \braket{\phi_{i}}{\Psi_n}$. This importance measure can be estimated within first-order MCPT, discussed in Sec. \ref{sec:mcpt}, starting from a multi-configurational reference state $\ket{\Psi_{\text{ref}}}$
\begin{equation}
 \kappa^{(1)}_{i} = \frac{\langle \phi_i \vert H \psiref}{E^{(0)}_i - E_{\text{ref}}^{(0)} } \, .
\end{equation}
The reference state provides an initial approximation of  $\ket{\Psi_n}$ within the reference space $\mathcal{M}_{\text{ref}}$ and $\kappa_{i}$ is used to assess {\it a priori} the importance of each basis state $\ket{\phi_i} \notin \mathcal{M}_{\text{ref}}$. Only states with importance measures above an importance threshold $\kappa_{\min}$ are included into the model space for the subsequent CI calculation. 

Since the eigenvalue problem is solved exactly in the importance-truncated model space, IT provides a variational approximation to the full solution. Moreover, the effect of discarded configurations can be estimated via the second-order MCPT energy correction \eqref{eq:mcpt2}. This together with extrapolations to vanishing importance thresholds $\kappa_{\min}\to 0$ can be used to reconstruct the energies in the full model space. As testified by the results already presented in Figs.~\ref{fig:MCPT_CFO}, \ref{fig:MCPT_spectra} and \ref{fig:BMBPT_comparison}, the IT concept has allowed one to push NCSM calculations beyond p-shell nuclei, up to neutron-rich oxygen isotopes. More details on importance-truncated CI approaches are discussed in Refs.~\cite{Roth2009,Stumpf2016}.

\subsection{New context: coupled-cluster method}

In the context of non-perturbative expansion methods, the general idea is to discard irrelevant entries of the largest mode-$n$ tensor that needs to be solved for and that drives the numerical scaling as well as the storage cost of the method. This is done by \emph{a priori} estimating the importance of each of its entries on the basis of a less costly many-body method, e.g. MBPT. While the underlying formalism is only sketched in this section, the reader is referred to Ref.~\cite{Tichai:2019ksh} for a detailed discussion.

\subsubsection{Wave-function ansatz}

To illustrate the concept, BMBPT is employed as a method to pre-process the cluster amplitudes constituting the unknown tensors to be solved for within the BCC method. This non-perturbative approach is based on the wave-function ansatz~\cite{Si15}
\begin{equation}
\label{e:bccwf}
| \Psi \rangle \equiv e^{\mathcal{T}} | \Phi \rangle \, ,
\end{equation}
where the connected quasiparticle cluster operator $\mathcal{T} \equiv \mathcal{T}_1 +\mathcal{T}_2 + \mathcal{T}_3 + \ldots$ is defined through
\begin{align}
\mathcal{T}_n &\equiv \frac{1}{n!}\displaystyle\sum_{k_1 \cdots k_{2n} }t^{2n0}_{k_1 \cdots k_{2n}} 
\beta^{\dagger}_{k_1}\cdots \beta^{\dagger}_{k_{2n}} \, , 
\end{align}
etc. The BCC amplitudes $t^{2n0}_{k_1 \cdots k_{2n}}$ constitute the unknown mode-$n$ tensors of present interest.

Instead of solving for all unknown cluster amplitudes, IT selects a subset of tensor entries that are determined non-perturbatively by solving BCC amplitude equations while residual corrections from the discarded tensor entries are treated in low-order BMBPT. Consequently, the ground-state energy $\mathcal{E}_0$ consists of a contribution from BCC in the IT subspace $\mathcal{E}_0^{(\text{IT})}$ and a residual BMBPT correction $\delta_0^{(\text{res})}$ from the complementary subspace
\begin{equation}
\label{e:bccenerPQ}
\mathcal{E}_{0} = \mathcal{E}_{0}^{(\text{IT})} + \delta_0^{(\text{res})} \, .
\end{equation}
Of course, the selection of the IT subspace  must be performed \emph{without the knowledge} of the full solution, which is where low-order BMBPT estimates enter into play.

\subsubsection{Importance-truncated tensor}

Considering the $n$-tuple BCC amplitude tensor $\mathcal{T}_n \equiv \{t^{2n0}_{k_1 \ldots k_{2n}}\}$, the corresponding \emph{importance-truncated} tensor based on the IT measure $\kappa^{(p)}(t^{2n0}_{k_1 \ldots k_{2n}})$ from BMBPT at order $p$ is obtained as
\begin{align}
\mathcal{T}^{(p)}_n(\kappa_\text{min}) \equiv \{t^{2n0}_{k_1 \ldots k_{2n}}  \,\, \text{such that}  \,\,  \kappa^{(p)} (t^{2n0}_{k_1 \ldots k_{2n}} ) \geq \kappa_\text{min} \} \, ,
\end{align}
where $\kappa_\text{min}$ defines the IT threshold. The original tensor is recovered
in the limit of $\kappa_\text{min} \rightarrow 0$, i.e.,
\begin{align}
\lim_{\Xmin{\kappa}\rightarrow 0} \mathcal{T}^{(p)}_n(\kappa_\text{min}) = \mathcal{T}_n \, .
\end{align}
With an importance-truncated tensor comes the associated \emph{data compression ratio} 
\begin{align}
R_c \equiv \frac{\text{\# elements of} \, \mathcal{T}_n}{\text{\# elements of} \, \mathcal{T}^{(p)}_n(\kappa_\text{min})} \, ,
\end{align}
which relates the initial amount of data to the compressed one resulting from the IT process. Whenever $R_c>1$, the compressed tensor requires less storage than the original one.

\subsubsection{Results}

Applying IT to $\mathcal{T}_3$, the goal is eventually (not done here) to solve BCCSDT equations for the retained entries and correct for the omitted ones in perturbation. The feasibility of the approach directly depends on the reduction offered by the IT for a desired accuracy given that a full BCCSDT calculation is currently undoable in realistic model spaces, even when employing an angular-momentum-coupled scheme. 

\begin{figure}[t!]
\centering
\includegraphics[width=0.49\textwidth]{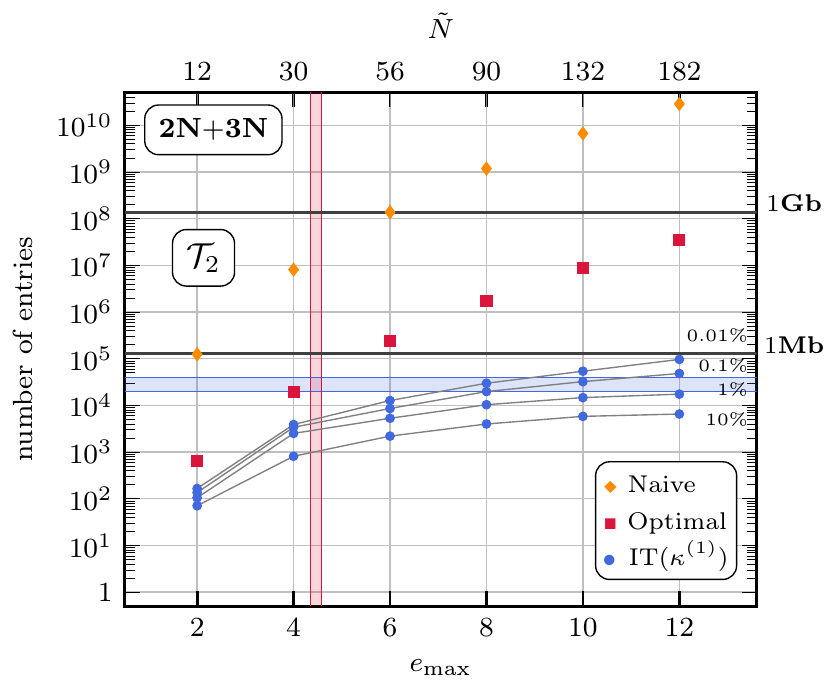}
\includegraphics[width=0.49\textwidth]{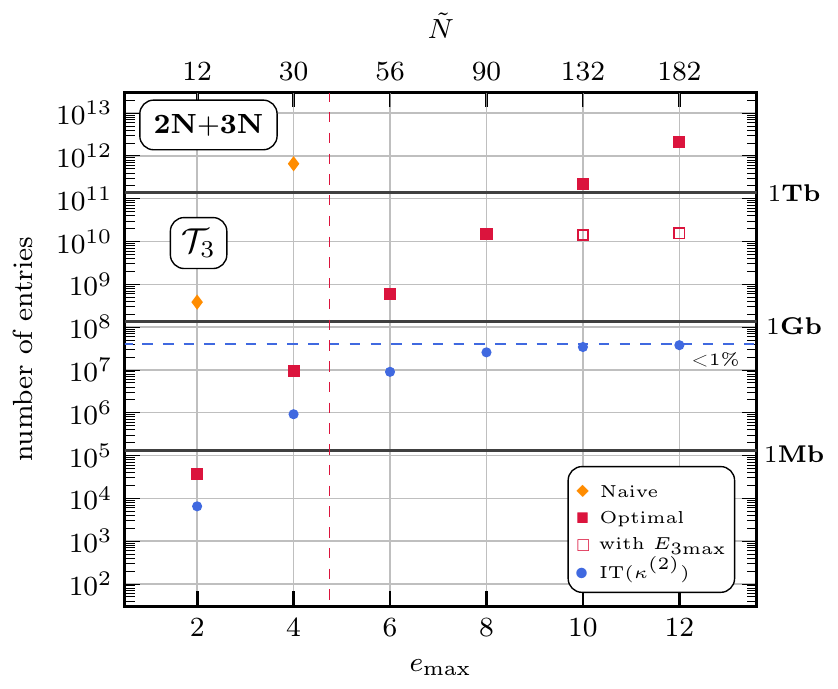}
\caption{Number of tensor entries of $\mathcal{T}_2$ amplitudes (left) and $\mathcal{T}_3$ amplitudes (right) as a function of model space using the Hamiltonian described in Sec.~\ref{sec:ham}. 
The oscillator frequency is given by $\hbar \omega=20\, \text{MeV}$.
Different plot markers indicate different storage scheme including all entries (\orangediamond), symmetry-allowed entries (\redsquare) and IT-compressed entries (\bluecircle). For $\Xmax{e} \geq 10$ an additional $E_{3\text{max}}$-truncation (\redsquareopen) for the $\mathcal{T}_3$ amplitudes is introduced.
All calculations employ a canonical HFB reference state.
Pictures taken from Ref.~\cite{Tichai:2019ksh}.
}
\label{fig:IT}
\end{figure}

Figure~\ref{fig:IT} displays the impact of the data compression obtained by applying IT for $^{18}$O as a function of the HO single-particle basis size. To provide a reference, the compression of $\mathcal{T}_2$ is illustrated first in the left panel before coming to the real challenge constituted by the treatment of $\mathcal{T}_3$. Three different storage schemes are displayed corresponding to a full storage of all tensor entries (diamonds), the tensor entries allowed by fundamental symmetries of the interaction (squares) and the IT-compressed entries from perturbative estimate (circle) for different values of $\kappa_\text{min}$. The percentage associated with the IT-data indicates the corresponding error on the second-order contribution to the ground-state energy
\begin{align}
\Delta \Omega^{(2)}_0 &= \frac{1}{4!}\sum_{ k_1 k_2 k_3 k_4 }  |t^{40(1)}_{k_1 k_2 k_3 k_4} |^2 E_{k_1 k_2 k_3 k_4} \, .
\end{align}
Results demonstrate that even for $\mathcal{T}_2$, i.e. a mode-4 tensor, the storage of all index combinations requires tremendous amounts of memory ($>200\,$GB per working copy) that is at the edge of supercomputing facilities\footnote{Solving non-linear problems requires the storage of various copies of the coupled-cluster amplitudes.}. Employing a symmetry-adapted storage scheme, several orders of magnitude can be saved. More importantly, discarding irrelevant tensor entries through IT additionally compresses the data  by three orders of magnitude at the price of inducing a systematic error of less than $1\%$, i.e., a few hundreds of keV, on the second-order correlation energy. Correspondingly, the storage of the  $\mathcal{T}_2$ amplitudes is lowered to less than $1\,$MB per working copy. One can assign an 'effective one-body dimension' to this number of tensor entries yielding $e_\text{max}^{\text{(eff)}}\leq 5$, i.e., converged BCCSD results could be obtained using an effective model space including less than six major shells.

While the storage of $\mathcal{T}_2$ amplitudes is actually within reach of state-of-the-art capacities, the extension to $\mathcal{T}_3$ amplitudes poses a severe computational problem. Performing the same analysis as above, even the symmetry-adapted tensor requires more than $100\,$TB of memory. Performing the IT-compression of $\mathcal{T}_3$, the corresponding fourth-order contribution to the ground-state energy
\begin{align}
\Delta \Omega^{[4_T]}_0 &= \sum_{ \substack{k_1 k_2 k_3 \\ k_4 k_5 k_6} }  |t^{60 (2)}_{k_1 k_2 k_3 k_4 k_5 k_6} |^2 E_{k_1 k_2 k_3 k_4 k_5 k_6} \, ,
\label{eq:Etriplesbody}
\end{align}
can be evaluated with $1\%$ accuracy while discarding $99.99\%$ of all $\mathcal{T}_3$ tensor entries. Because triples correction to the ground-state energy first appear at fourth order, a one-percent relative error on $\Delta \Omega^{[4_T]}_0$ relates to an absolute error of a few keV, which is negligible in mid-mass \emph{ab initio} studies. 

Eventually, the use of IT shifts the boundary of what is computationally unfeasible such that, in the following years, the implementation of IT-BCCSDT defines an ambitious and yet reachable goal.

\section{Basis optimization}
\label{sec:natorbs}

The present section discusses how MBPT can be used as an auxiliary tool to construct an alternative single-particle basis accelerating the convergence of non-perturbative many-body methods as compared to commonly employed bases. 

\subsection{Rationale}

In nuclear structure calculations, tensors defining $H$ are typically expressed in the eigenbasis of the spherical HO Hamiltonian. Next, a self-consistent mean-field, e.g. HF, calculation is typically performed to generate the reference state such that the beyond-mean-field step and the associated tensors are transformed into the corresponding mean-field, e.g. HF, one-body basis.

While the use of HO orbitals authorizes the rigorous center-of-mass factorization in NCSM calculations, they are plagued at long distances with a pathological Gaussian falloff and a strong dependence of the many-body results on the width of the confining HO potential. Contrarily, HF orbitals display a proper exponential falloff but may induce a sizeable center-of-mass contamination and a slower convergence in large-scale NCSM diagonalizations compared to HO orbitals.

\subsection{Natural orbitals}

Naturally, one wonders about the existence of a single-particle basis that admit only little sensitivity to the oscillator frequency while maintaining rapid model space convergence. Because the exact one-body density matrix
\begin{align}
\rho_{pq} = \frac{\la \Psi | c^\dagger_p c_q  | \Psi \ra}{\la \Psi | \Psi \ra} \, ,
\end{align}
contains information about the fully correlated wave function, its eigenbasis, defining so-called \emph{natural orbitals}, is expected to deliver an optimal choice. 

This expectation was confirmed by employing the one-body density matrix obtained from a large-scale NCSM calculation~\cite{Constantinou2017}. For mid-mass nuclei, this strategy has the downside to require a full CI solution to obtain the optimized single-particle basis. However, as already observed several decades ago in quantum chemistry calculations~\cite{Hay1973,Siu1974}, \emph{approximate} natural orbitals perform surprisingly well in applications. Instead of using the exact CI wave function, a correlated density matrix is constructed within a polynomially scaling expansion method, MBPT providing a particularly simple example to do so~\cite{Tichai2019a}. Similar approaches can be followed by using the $\Lambda$-extension of CC theory or by diagonalizing the dressed one-body propagator in SCGF theory. Note, however, that the construction of an auxiliary basis already involves a computationally non-trivial solution within a non-perturbative many-body approach.

\subsection{Density matrix in closed-shell MBPT}

Starting from Eq.~\eqref{eq:powerseriesWF} and following Ref.~\cite{Strayer1973}, the one-body density matrix up to second order in the residual interaction can be written as~\cite{Tichai2019a}
\begin{align}
\label{eq:diags}
\rho = \rho^{(00)} + \rho^{(02)} + \rho^{(20)} + \rho^{(11)} + \mathcal{O}(\lambda^3)\, , 
\end{align}
where 
\begin{align}
\rho^{(00)}_{pq} \equiv \frac{\la \Phi | c^\dagger_p c_q | \Phi \ra }{\la\Phi | \Phi \ra}
\end{align}
denotes the zeroth-order HF density matrix whereas
\begin{subequations}
\begin{align}
\rho^{(02)}_{pq} &\equiv \la \Phi^{(0)} \vert\,  c_p^\dagger c_q\, \vert \Phi^{(2)} \ra = \rho^{(20)\ast}_{qp} , \\
\rho^{(11)}_{pq} &\equiv \la \Phi^{(1)} \vert\, c_p^\dagger c_q\, \vert \Phi^{(1)} \ra \, ,
\end{align}
\end{subequations}
denote its MBPT corrections whose matrix elements are expressed in the HF basis. Explicit expressions for the individual corrections to the mean-field density in terms of two-body matrix elements can be found in Refs.~\cite{Strayer1973,Tichai2019a}.

\begin{figure}[t!]
\centering
\includegraphics[width=0.7\textwidth]{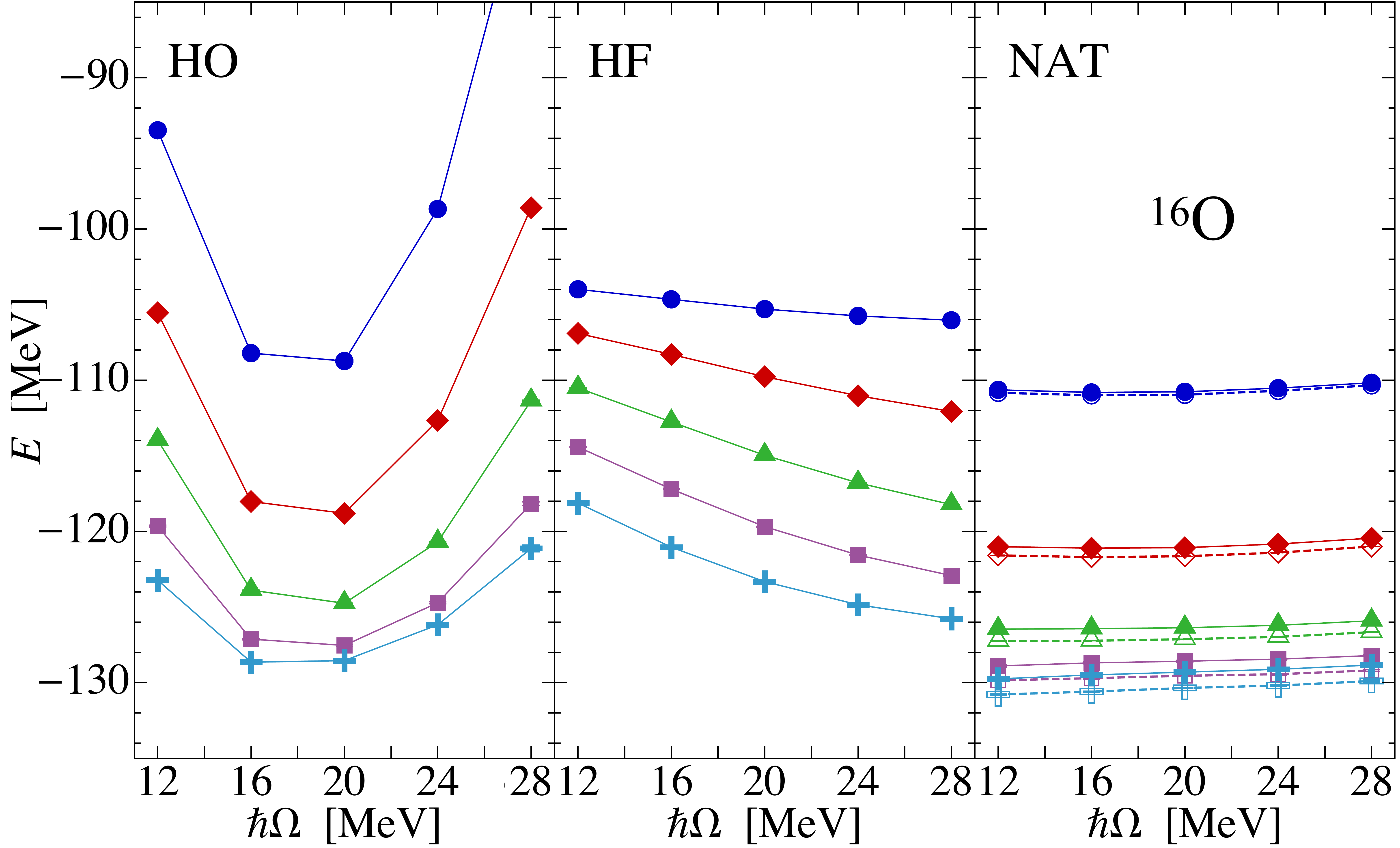}
\caption{NCSM ground-state energy of $^{16}$O  as a function of the frequency $\hbar \omega$ (denoted as $\hbar \Omega$ on the figure) of the underling HO basis for $N_{\max}=2,4,6,8,10$ spaces. The NCSM diagonalization itself is performed either in the initial HO basis (left), in the HF basis (center) or in the natural-orbital basis from second-order HF-MBPT (right). The Hamiltonian described in Sec.~\ref{sec:ham} is employed. Figure taken from Ref.~\cite{Tichai2019a}.
}
\label{fig:NO_O16}
\end{figure}

\subsection{Results}

The impact of using natural orbitals is now gauged by performing a systematic variation of the underlying single-particle basis in NCSM calculations. Figure~\ref{fig:NO_O16} displays  $^{16}$O ground-state energy as a function of the frequency characterizing the underling HO basis for various values of $N_{\max}$. Results obtained from the NCSM diagonalization performed in the HO  and HF bases as well as in the natural-orbital basis extracted from second-order HF-MBPT are displayed.

The left panel shows the typical strong dependence of NCSM results on the underlying frequency when using HO one-body basis states, thus yielding a pronounced parabolic shape around the variational minimum for all values of $N_\text{max}$. While the use of HF orbitals (middle panel) lowers the sensitivity to $\hbar \omega$ the model space convergence is not improved and the ground-state energy has a linear falloff at higher frequencies. Finally, the NCSM ground-state energy obtained using the natural-orbital basis (right panel) is almost independent of the oscillator frequency and displays a faster model-space convergence as a function of $N_\text{max}$. Similar conclusions hold for other observables, e.g., charge radii, low-lying excitation energies and electro-magnetic transition strengths, as discussed in Ref.~\cite{Tichai2019a}.

Eventually, the use of one-body density matrices from low-order MBPT to construct of an optimal computational basis is a simple yet powerful tool to account for high-lying particle-hole excitations that are otherwise hard to grasp in configuration-driven methods like NCSM.

\section{Testing novel $\chEFT$ nuclear Hamiltonians}
\label{sec:newham}

The computational efficiency of perturbative approaches makes them ideally suited for survey calculations over a large range of nuclei. A specific scenario for such survey calculations is the development of novel chiral EFT interactions, where large numbers of many-body calculations are necessary to constrain or validate the choice of low-energy constants as well as to characterize the effect of different regulator formulations. Recently, a new family of consistent chiral 2N plus 3N interactions up to N3LO based on non-local regulators was developped \cite{Huether2019}. This new set of interactions is able to simultaneously reproduce ground-state energies and charge radii up into the medium-mass regime. Moreover, the full sequence of chiral orders from LO to N3LO is available for different cutoff values, which facilitates a rigorous uncertainty quantification. In the applications discussed below, the new interactions are consistently SRG-evolved in the 2N and 3N sectors down to $\alpha = 0.04$ fm$^4$ and tested through non-expensive perturbative methods over a larger set of nuclei than those addressed in Ref.~\cite{Huether2019}.

\subsection{NCSM-PT}

\begin{figure}[t!]
\centering
\includegraphics[width=0.75\textwidth]{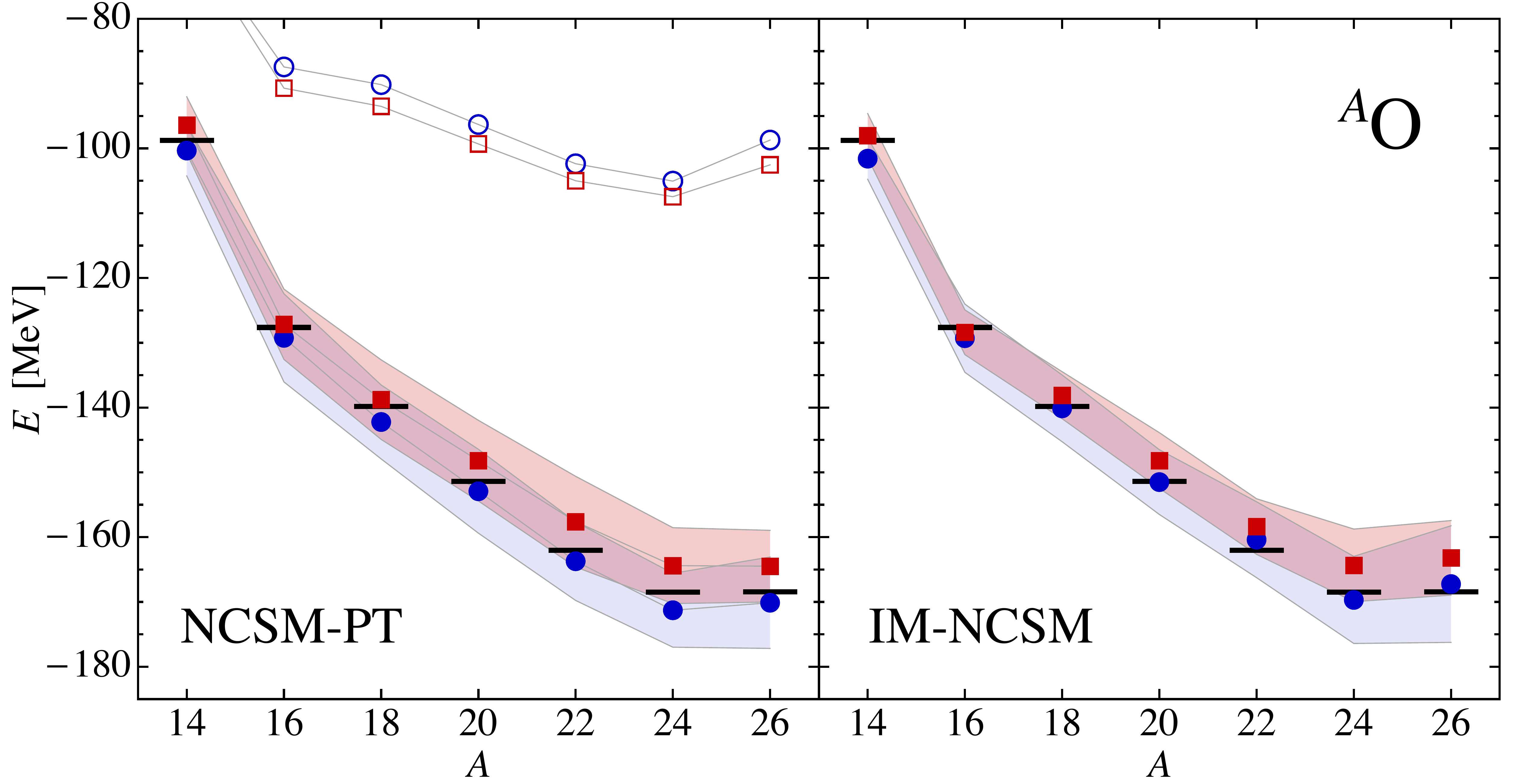}
\caption{Even oxygen isotopes ground-state energies from reference (empty symbols) and second-order (full symbols) NCSM-PT with $N_{\max}^{\text{ref}}=2$ (left panel) and from IM-NCSM (right panel). The consistent non-local 2N plus 3N chiral interactions at N$^2$LO (\bluecircle) and N$^3$LO (\redsquare) presented in Ref. \cite{Huether2019} are employed with a cutoff $\Lambda=500$ MeV. The uncertainty bands represent both many-body and interaction uncertainties (see text). Both many-body calculations make use a natural-orbital basis obtained in 13 oscillator shells with $\hbar \omega=20\, \text{MeV}$.}
\label{fig:MCPT_hfnat}
\end{figure}

First, NCSM-PT is employed to calculate ground-state energies along the oxygen chain using interactions from Ref. \cite{Huether2019}. In Fig. \ref{fig:MCPT_hfnat}, results obtained from N2LO and N3LO interactions are compared to experiment including a quantification of systematic theoretical uncertainties resulting from both the many-body and interaction truncations. The benchmarks discussed in Sec. \ref{sec:mcpt} showed that calculations with the $N_{\max}^{\text{ref}}=2$ typically provide a good agreement with non-perturbative methods and that ground-state energies for larger reference spaces typically fall between $N_{\max}^{\text{ref}}=0$ and $2$ results. Therefore, the difference between these two calculations offers a convenient way to estimate many-body uncertainties in NCSM-PT calculations. Uncertainties resulting from truncating the chiral expansion of nuclear interactions are estimated from the order-by-order variation of the energies using the prescription discussed in Refs.~\cite{Epelbaum2015,LENPIC2016,LENPIC2018}. The interaction-induced uncertainty of an observable $X_{\text{N3LO}}$ at order N3LO, e.g., is given by  $\max( Q\, | X_{\text{N3LO}} - X_{\text{N2LO}} |, Q^2\, | X_{\text{N2LO}} - X_{\text{NLO}} |, Q^3\,| X_{\text{NLO}} - X_{\text{LO}} |, Q^5\,| X_{\text{LO}} | )$ with the expansion parameter $Q$ estimated as the ratio of a typical momentum scale characterizing mid-mass systems over the breakdown scale, which results in $Q\approx1/3$~\cite{LENPIC2018}. The uncertainty band shown in Fig. \ref{fig:MCPT_hfnat} combines both uncertainties, where many-body uncertainties are typically larger than interaction uncertainties.

The NCSM-PT ground-state energies can be compared with results from the in-medium no-core shell model (IM-NCSM) displayed on the right-hand side of Fig. \ref{fig:MCPT_hfnat}. The non-perturbative IM-NCSM method combines a multi-reference in-medium SRG transformation of the Hamiltonian that decouples a small NCSM model space, with a subsequent NCSM calculation using the transformed Hamiltonian \cite{Geb17}. The associated uncertainty bands include once again both many-body and interaction uncertainties. The comparison reveals an excellent agreement between NCSM-PT and IM-NCSM, which themselves agree with experiment within estimated uncertainties.

\subsection{BMBPT}
\label{loworderBMBPTresults2}

In addition to low-order NCSM-PT benchmarks in oxygen isotopes, the novel set of interactions is tested on a large set of mid-mass semi-magic nuclei up to third-order in BMBPT\footnote{As compared to Fig.~\ref{fig:BMBPT_detail}, third-order are presently corrected for the particle-number contamination via the so-called {\it a posteriori} correction studied and validated in Ref.~\cite{Demol:2020}.}. 

To estimate the many-body uncertainty arising from the finite truncation of the BMBPT expansion a nucleus-independent error of $3\%$ is assumed. A recent study of high-order BMBPT calculations revealed that the third-order partial sum typically induces a many-body error of about $2\%$~\cite{Demol:2020}. Because the employed Hamiltonian was softer and further SRG-evolved to lower resolution scale, a more conservative $3\%$-error in the current setting seems plausible. The overall uncertainty is then given by summing this many-body error and the interaction uncertainty stipulated before. The additional uncertainty associated with the  model space truncation induced by the finite single-particle basis is presently not incorporated. Clearly, the construction of reasonable models of many-body uncertainties is highly non-trivial and employing simplistic parametric models, e.g., geometric sums for perturbation series, can lead to wrong estimates if the perturbation series does not obey the underlying assumptions.

\begin{figure}[t!]
\centering
\includegraphics[width=0.75\textwidth]{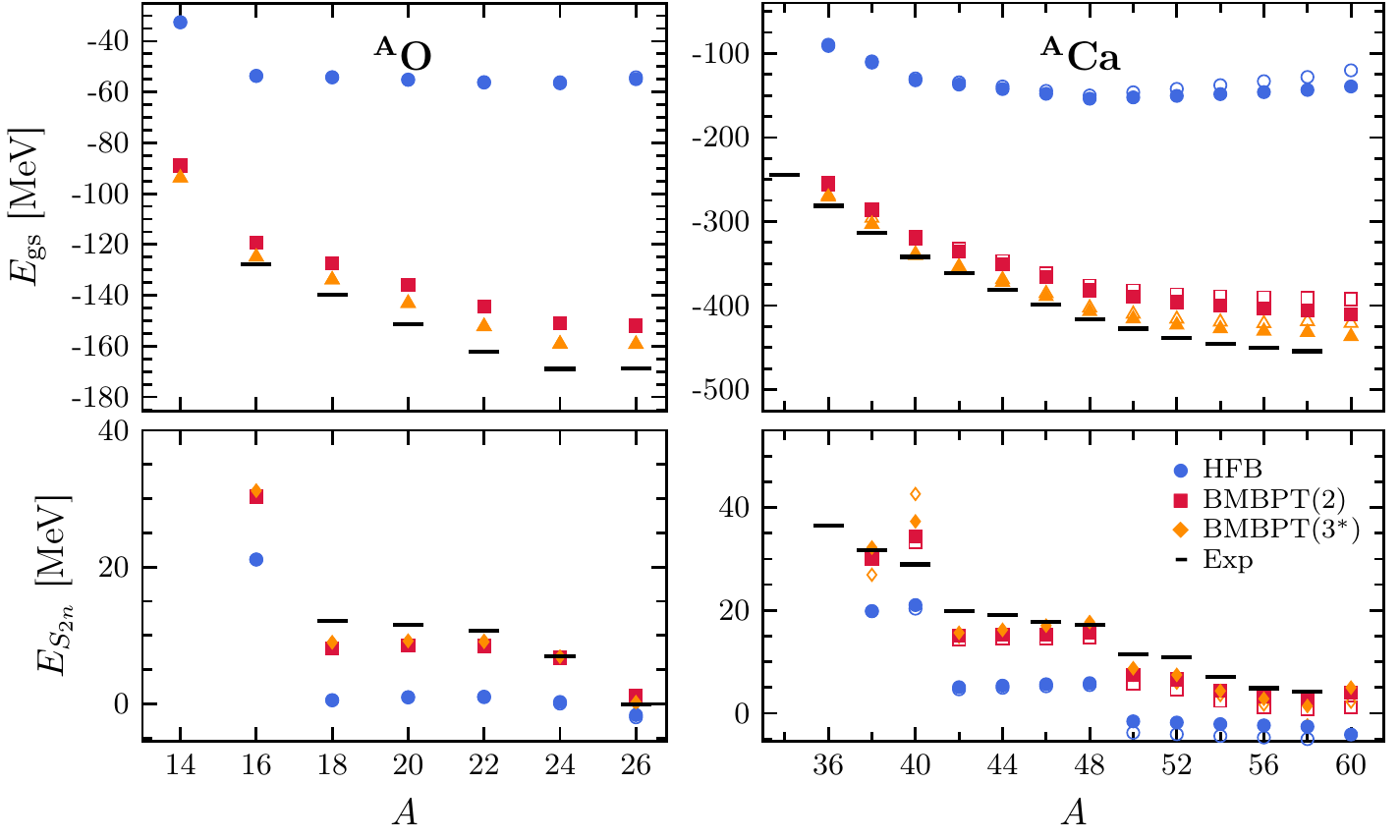}
\caption{BMBPT systematics along O (left) and Ca (right) isotopic chains using the non-local chiral two- plus three-nucleon interactions at N$^3$LO presented in Ref. \cite{Huether2019}. Top panel: absolute binding energies. Bottom panel: two-neutron separation energies. Plot markers correspond to HFB (\bluecircle), second-order BMBPT (\redsquare) and third-order BMBPT (\orangediamond). All calculations are performed using 15 oscillator shells and an oscillator frequency of $\hbar \omega=20\, \text{MeV}$. The single-particle orbital angular-momentum quantum number was limited to $l\leq 10$. Open and closed symbols correspond to different truncations $E_{3\text{max}}=14,16$, respectively.}
\label{fig:BMBPT_new}
\end{figure}

In Fig.~\ref{fig:BMBPT_new}, ground-state energies of oxygen and calcium isotopes are displayed for two different values of $E_{3\text{max}}=14,16$. The first striking result is that the overbinding obtained in Ca isotopes with the 'standard' Hamiltonian (see Fig.~\ref{fig:BMBPT_comparison}) is resolved, i.e. the systematic trend throughout O and Ca isotopes is now consistent with experimental data. In particular, low-order BMBPT results provide a much better reproduction of experimental ground-state energies of very neutron-rich Ca isotopes than with the 'standard' Hamiltonian for which the degrading increases systematically with neutron number. 

Going in more details, one observes that the reference HFB energy, i.e., the lowest order in the BMBPT expansion, accounts for much less binding (i.e. less than half of the total binding energy) than with the 'standard' Hamiltonian, thus indicating a 'harder' interaction\footnote{This feature reflects both the different characters of the chiral Hamiltonians themselves and the fact that the 'standard' Hamiltonian was SRG-evolved down to  $\alpha = 0.08$ fm$^4$ instead of  $\alpha = 0.04$ fm$^4$ for the new ones.}. Moreover, the third-order correction is typically larger in the present case. While the the first three BMBPT orders are systematically and strongly suppressed, this indicates that the new interactions are indeed less perturbative. One can thus expect higher orders to contribute non-negligibly.

Focusing now on two-neutron separation energies, the end results are very satisfactory and of similar or even greater quality than with the 'standard' Hamiltonian. Interestingly, HFB results are further away in the present case, indicating that not only absolute values but also the trend with neutron number is different. Still, second and third order contributions consistently compensate for this apparent, but in fact fictitious, degrading at the HFB level. One eventually observes that results are still less accurate near major shell closures, which is consistent with the expectation that restoring particle-number symmetry through PBMBPT~\cite{Duguet:2015yle} will have the largest impact near shell closures.  

While the results in oxygen isotopes are virtually identical for both $E_{3\text{max}}$ values, the truncation in $E_{3\text{max}}$ plays an increasingly important role in  neutron-rich calcium isotopes, i.e. ground-state energies beyond $A=50$ obtained for $E_{3\text{max}}=14$ and $16$ start deviating more strongly, pointing to the importance of truncated three-body matrix elements as already shown in Ref.~\cite{Huether2019}. This constitutes a technical challenge for the future when \textit{ab initio} calculations move to even heavier and more neutron-rich nuclei. 

Clearly, the above features points toward a systematically improved quality of the novel set of  interactions compared to prior generations of chiral Hamiltonians that tend to fail in mid-mass systems. In the future, this statement will be further benchmarked by validating the consistency of low-lying spectroscopy and charge radii with experimental data in the mid-mass region.

\begin{figure}[t!]
\centering
\includegraphics[width=0.75\textwidth]{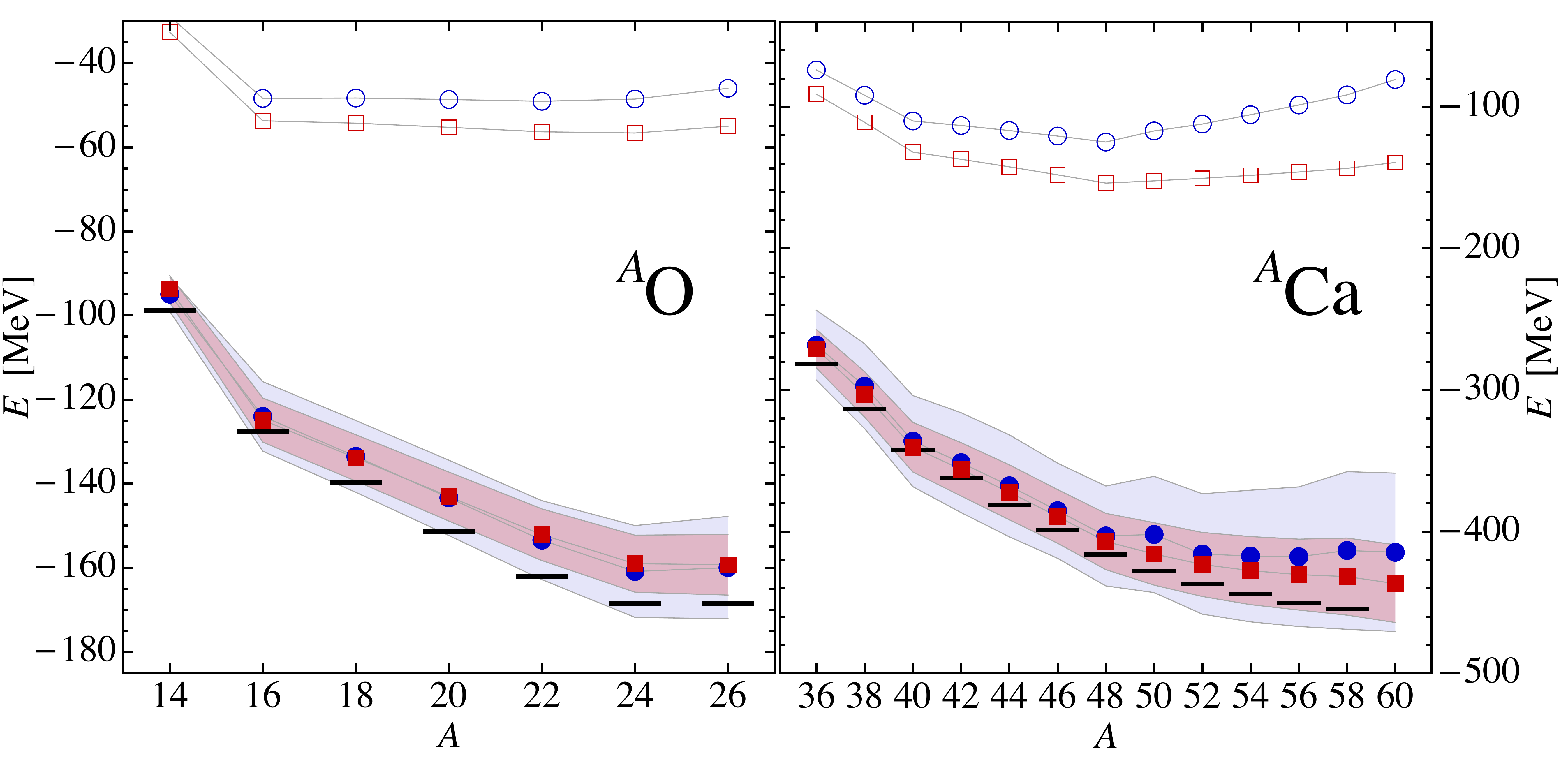}
\caption{BMBPT ground-state energies for oxygen and calcium isotopes at third order of the perturbative expansion using consistent non-local chiral two- plus three-nucleon interactions at N$^2$LO (blue) and N$^3$LO (red). Open symbols show the corresponding HFB ground-state energies. The uncertainty bands represent both, many-body and interaction uncertainties (see text). All calculations are performed using 15 oscillator shells and an oscillator frequency of $\hbar \omega=20\, \text{MeV}$. The single-particle orbital angular-momentum quantum number was limited to $l\leq 10$.}
\label{fig:BMBPT_UQ}
\end{figure}

\section{Conclusions and Outlook}
\label{sec:conc}

The present paper reviewed the status of MBPT techniques in the field of \emph{ab initio}  many-body calculations of finite nuclei. After discussing formal properties of the power-series ansatz, the main goal was to provide an in-depth understanding of the MBPT expansion and its range of applicability. Most importantly, two ways of extending Slater-determinant-based MBPT towards open-shell systems were discussed. Multi-configurational perturbation theory and Bogoliubov MBPT were shown to provide complementary ways to tackle the particle-hole degeneracy and the associated strong correlations in open-shell nuclei.

Each of the subparts dedicated to the various flavors of many-body perturbation theory displayed results of large-scale calculations touching upon the very frontier of current \emph{ab initio} studies dedicated to mid-mass (open-shell) nuclei. While highly accurate non-perturbative alternatives are available, the formal and computational simplicity is at the very heart of perturbative approaches. Indeed, these features enable large surveys along complete isotopic chains at a small fraction of the computational cost necessary to utilize more sophisticated approaches. In view of the increasing activity related to the construction of improved generations of nuclear Hamiltonians, simple and yet accurate many-body schemes constitute a very valuable tool to quickly characterize their overall performance.

Furthermore, MBPT provides not only a standalone many-body framework, but can also be used as an inexpensive pre-processing tool to accelerate non-perturbative methods. Two facets discussed in this work are data compression from MBPT-based importance truncation and basis optimization from low-order MBPT density matrices. Both tools have been used with great success to either tame the curse of dimensionality or cure pathological defects in the computational basis at low computational price.

This (by far not exhaustive) list of MBPT-related applications in nuclear theory highlights the start of a \emph{renaissance of perturbation theory} fostered by renormalization group techniques used to soften nuclear Hamiltonians. In parallel, the deepened understanding of infrared-divergences, their origin in open-shell systems and systematic ways to overcome them will hopefully help cure the (unfortunately still present) disfavour against MBPT for nuclear structure research.

Future developments will expand many of the ideas laid out in this work. Better control on uncertainties in the nuclear Hamiltonian pave the way for targeting more exotic systems in the upcoming years. In particular, a proper account of static correlations associated with nuclear deformation has been identified as a critical goal by various groups worldwide. Consequently, the implementation of frameworks employing a systematic breaking (and restoration) of $SU(2)$ symmetry will be of high relevance. However, due to the increasing number of basis states these efforts will require significant computational resources as well as extensive formal developments. To control the increase of computational requirements (memory and runtime), so-called \emph{tensor factorization techniques} have been proposed recently where high-mode tensors are decomposed into sums of product of lower-rank ones~\cite{Tichai:2018eem,Tichai:2019ksh}. While initial proof-of-principle applications have demonstrated the high potential, extensive additional research is required in this direction.

On a short timescale the benchmarking of importance truncation and natural orbitals in non-perturbative medium-mass applications will deepen the understanding of the role of the computational basis and distribution of correlations in Hilbert space. Furthermore, by employing data compression tools will greatly help relax the many-body truncation truncation at play such that first-principle calculations will witness a level of accuracy that has been out of reach so far.

\section*{Funding}
This publication is based on work supported in part by the framework of the Espace de Structure et de r\'eactions Nucl\'eaires Th\'eorique (ESNT) at CEA, the  Deutsche  Forschungsgemeinschaft  (DFG,  German Research Foundation) – Projektnummer 279384907 – SFB 1245, and the BMBF through contract 05P18RDFN1 and Verbundprojekt 05P2018 (ErUM-FSP T07).

\section*{Acknowledgements}
We would like to thank M. Frosini for generating the data used in Fig.~\ref{fig:divergence}.
Parts of this work were performed at the Lichtenberg cluster at the computing center of the TU Darmstadt and the JURECA supercomputing facility at Forschungszentrum J\"ulich.

\bibliographystyle{frontiersinHLTH&FPHY} 
\bibliography{abbrev,bib_clean}

\end{document}